\documentclass[10pt,aps,prd,nofootinbib,superscriptaddress,twocolumn,preprintnumbers,balancelastpage,groupedaddress]{revtex4-1}

\usepackage{graphicx,epsfig,psfrag,amssymb,hyperref}
\usepackage{multirow}
\usepackage{color,graphicx,epsfig,psfrag,amsmath,empheq}
\usepackage{bm}
\usepackage{mathrsfs,amsfonts,color}
\usepackage{slashed}
\usepackage{hepunits}
\usepackage{color}
\usepackage{enumitem}
\usepackage{comment}
\definecolor{light-gray}{gray}{0.9}

\newcommand{\be}{\begin{equation}}
\newcommand{\ee}{\end{equation}}
\newcommand{\beq}{\begin{equation}}
\newcommand{\eeq}{\end{equation}}
\newcommand{\nl}{\nonumber \\}

\newcommand{\E}{\boldsymbol{E}}
\newcommand{\Ep}{\boldsymbol{E}^\prime}
\newcommand{\B}{\boldsymbol{B}}
\newcommand{\Bp}{\boldsymbol{B}^\prime}

\newcommand{\x}{\chi}
\newcommand{\Ap}{A^\prime}
\newcommand{\mAp}{m_{A^\prime}}
\newcommand{\Lag}{\mathscr{L}}
\newcommand{\eps}{\epsilon}
\newcommand{\p}{\prime}

\newcommand{\km}{\text{km}}
\newcommand{\mK}{\text{mK}}
\newcommand{\s}{\text{s}}
\newcommand{\kV}{\text{kV}}

\newcommand{\lap}{\nabla}
\newcommand{\rhodm}{\rho_{_{\text{DM}}}}
\newcommand{\tint}{t_\text{int}}
\newcommand{\order}[1]{\mathcal{O}{(#1)}}
\newcommand{\qeff}{q_\text{eff}}

\newcommand{\jp}{\boldsymbol{j}^\prime}
\newcommand{\jx}{\boldsymbol{j}_\x}
\newcommand{\jv}{\boldsymbol{j}}
\newcommand{\G}{\boldsymbol{G}}

\newcommand{\xv}{{\bf x}}
\newcommand{\yv}{{\bf y}}
\newcommand{\vv}{{\bf v}}

\newcommand{\Evis}{\boldsymbol{E}_\text{vis}}
\newcommand{\Bvis}{\boldsymbol{B}_\text{vis}}
\newcommand{\tp}{t^\prime}
\newcommand{\tpp}{t^{\prime \prime}}
\newcommand{\defl}{\text{def}}
\newcommand{\eqbox}[1]{\text{\fcolorbox{light-gray}{light-gray}{$#1$}}}

\definecolor{colorRTD}{rgb}{.2,.2,.7}

\begin{document}

\title{Direct Deflection of Particle Dark Matter}

\author{Asher Berlin}
\author{Raffaele Tito D'Agnolo}
\author{Sebastian A. R. Ellis}
\author{Philip Schuster}
\author{Natalia Toro}
\affiliation{SLAC National Accelerator Laboratory, 2575 Sand Hill Road, Menlo Park, CA 94025, USA}

\date{\today}

\begin{abstract}

We propose a new strategy to directly detect light particle dark matter that has long-ranged interactions with ordinary matter. 
The approach involves distorting the local flow of dark matter with time-varying fields and measuring these distortions with shielded resonant detectors. 
We apply this idea to sub-MeV dark matter particles with very small electric charges or coupled to a light vector mediator, including the freeze-in parameter space targeted by low mass direct detection efforts. This approach can probe dark matter masses ranging from 10 MeV to below a meV, extending beyond the capabilities of existing and proposed direct detection experiments. 

\end{abstract}

\maketitle

Dark matter (DM) constitutes the majority of matter in the universe, but an understanding of its nature and interactions remains elusive. The landscape of viable candidates is vast, ranging in mass from $\sim 10^{-22} \ \eV$ to superplanetary scales, with a broad variety of possible interactions.  Moreover, DM may have particle-like (particle-number-conserving) or field-like (linear-in-DM-field and hence, particle-number-violating) interactions. These two classes of interaction can occur in overlapping mass ranges, but give rise to very different phenomenology.  This wide range of possibilities motivates employing a varied set of detection strategies (see Ref.~\cite{Battaglieri:2017aum} for an overview). 

Searches for sub-eV DM often assume interactions linear in the DM field and exploit the semiclassical properties of the coherent DM field for detection.  Furthermore, these searches often rely on detectors that are resonantly matched to the angular frequency of the oscillating DM field (set by the DM constituent mass).  Searches for heavier field-like DM, up to $\keV$-scale masses, exploit the absorption of DM particles by electrons or nuclei through these single-field couplings. On the other hand, detection strategies for DM with an exact (or approximate) DM-particle-number symmetry rely on single-particle scattering reactions. These setups can only observe DM heavy enough that a single scattering event transfers an observably large energy to the detector. As a result, such experiments are most sensitive in the $10 \ \GeV-\TeV$ mass-range, but recent technological advances are enabling sensitivity to $\MeV$-scale DM scattering and may eventually reach $\keV$ mass thresholds~\cite{Battaglieri:2017aum}.  

In this work, we propose a new  approach (``direct deflection'') to search for sub-MeV particle-like DM that exploits long-range interactions between DM and ordinary matter, as expected in prominent freeze-in scenarios and any model of DM with an ultralight force mediator. The approach is based on inducing collective effects in the DM fluid on detector length-scales that can leave a measurable trace in resonant detectors. In the \emph{collectiveness} of the effect, and in its enabling technologies, direct deflection is reminiscent of light field-like DM searches. However, the signal does not depend on DM behaving like a classical \emph{field} (large occupation number per cubic de Broglie wavelength, with linear coupling to matter), but instead on the far laxer condition that it behaves like a classical \emph{fluid} (large number of DM particles in the volume of the apparatus, with number-conserving interactions).  In models that can be probed by both direct deflection and traditional direct detection experiments, the deflection approach has a parametric advantage at low DM masses, where the low energy deposited by single-particle scattering becomes difficult to measure but collective effects of the DM fluid are enhanced. 

The general direct deflection concept can be realized in multiple ways, potentially targeting a variety of motivated DM models. Here, we focus on a concrete setup that is sensitive to DM particles coupled to a light kinetically-mixed vector mediator.  One motivation for exploring such models is  
so-called ``freeze-in'' DM that arises through very feeble interactions between DM and Standard Model (SM) matter~\cite{Hall:2009bx}.  Notably, for $\keV-\MeV$ DM masses, freeze-in through a very light mediator is among the very few models known to have a viable and predictive cosmology while being plausibly explorable with terrestrial experiments.  This scenario has become an important benchmark for low-mass direct detection experiments~\cite{Battaglieri:2017aum}, and exhibits terrestrial phenomenology akin to millicharged particles.  

With this motivation in mind, our proposal is to induce and subsequently detect oscillating effective DM \emph{millicharge} or \emph{millicurrent} densities, using large driven electromagnetic fields and well-shielded resonant detectors, similar to ``light-shining-through-a-wall"-type experiments~\cite{Jaeckel:2007ch,Betz:2013dza,Graham:2014sha}. Unlike such experiments, however, our setup does not rely on the production of new light states. Instead, ambient DM from the Milky Way halo passing through an oscillating electromagnetic field is deflected, setting up propagating waves of DM millicharge and millicurrent. These DM waves can penetrate a downstream electromagnetic shield by virtue of the tiny DM coupling (see, e.g., Fig.~\ref{fig:reach}), establishing small oscillating electric and magnetic fields that can be measured with a resonator coupled to a sensitive magnetometer. A schematic illustration of this ``wind-blowing-through-a-wall" apparatus is shown in Fig.~\ref{fig:setup}. This technique is based on tested technology, complements and competes with other direct detection proposals in the $\keV-\MeV$ mass-range, and is sensitive to much smaller masses, going beyond current astrophysical constraints for DM lighter than a keV.  The anticipated reach of various experimental configurations is shown in Fig.~\ref{fig:reach}. 

\emph{Millicharged and Millicharge-like Dark Matter.} 
To illustrate our idea, we consider the scenario in which DM, denoted by $\x$, couples to standard electromagnetism with an effective charge $\qeff \ll 1$. This is often referred to as ``millicharged'' DM, and in its simplest incarnation requires no new particles beyond the DM itself. A natural way for such effective models to arise is when the DM is charged under a hidden sector gauge boson $\Ap_\mu$ (a ``dark photon'') that kinetically mixes with the SM photon~\cite{Holdom:1985ag},
\be
\label{eq:kinmix}
\Lag \supset \frac{\eps}{2} \, F_{\mu \nu} \, F^{\p \mu \nu} + \frac{1}{2} \, \mAp^2 \, A_\mu^{\p 2}
\, ,
\ee
where $\mAp$ is the dark photon mass and the dimensionless coupling, $\eps \ll 1$, controls the strength of kinetic mixing.
DM coupled to a massless dark photon ($m_{\Ap}=0$) induces an effective millicharge 
\be
\label{eq:qeff}
\qeff = \eps \, e^\p / e
\, ,
\ee
in addition to DM self-interactions controlled by $e^\p$, where $e$ is the SM electric charge and $e^\p$ is the dark photon gauge coupling.  For non-zero $m_{\Ap}$, DM interactions with SM matter are millicharge-like over distance scales $\lesssim m_{\Ap}^{-1}$ and exponentially screened at larger distances. We consider an experimental apparatus localized to $\order{\text{meter}}$-scale distances, for which $\mAp = 0$ and $\mAp \lesssim \text{meter}^{-1} \sim 10^{-7} \ \eV$ are qualitatively indistinguishable. We focus on the massless case for simplicity and discuss finite-mass corrections in the Supplementary Material. 

\begin{figure}[t!]
\includegraphics[width = 0.48\textwidth]{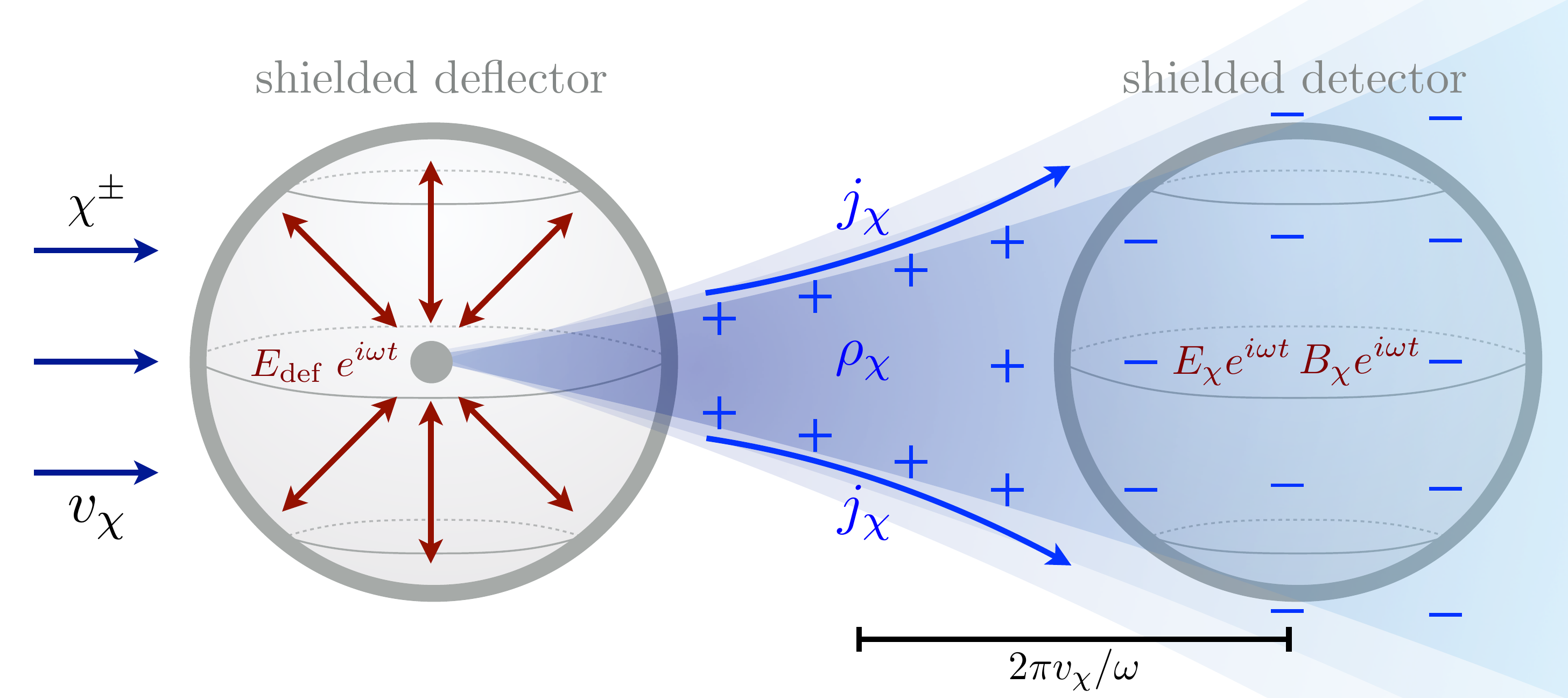}
\vspace{-0.7cm}
\caption{Schematic of the experimental concept. Dark matter passing through an oscillating electric field is deflected, setting up propagating waves of alternating dark matter millicharge, $\rho_\x$, and millicurrent, $j_\chi$, densities. Dark matter waves flow unimpeded through an electromagnetic shield, creating small electromagnetic fields that can be measured with a resonant LC circuit inductively coupled to a magnetometer (not depicted) inside the shielded detector region.
}
\label{fig:setup}
\vspace{-0.5cm}
\end{figure}
\begin{figure}[t!]
\includegraphics[width = 0.5\textwidth]{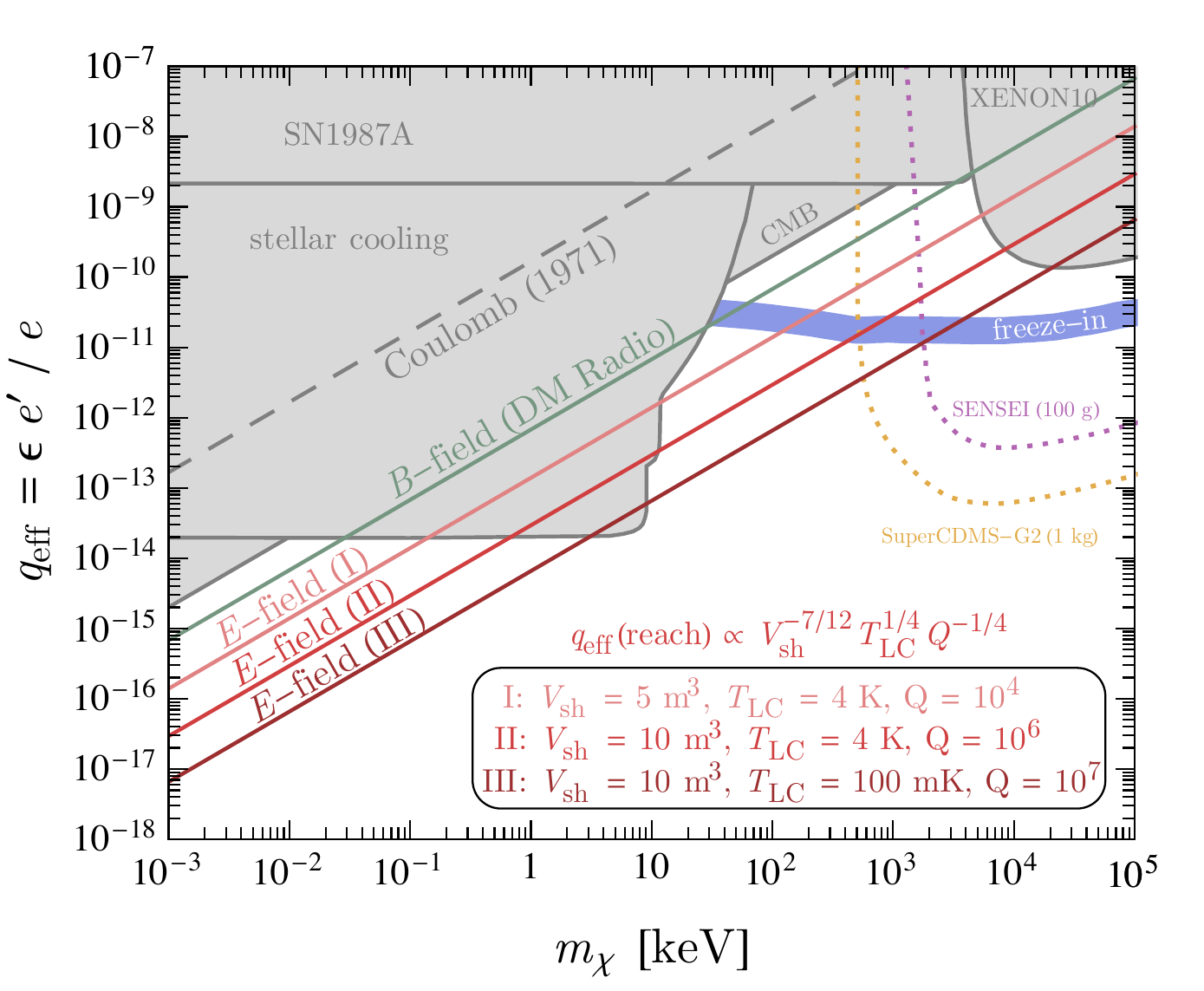}
\vspace{-0.7cm}
\caption{The anticipated reach to millicharged dark matter in the $\qeff - m_\x$ plane for various experimental configurations of our setup at 90\% C.L., compared to existing constraints (shaded gray). In all cases, we assume a year of integration time, a spatially-averaged field-strength of $\langle E_\defl \rangle = 10 \ \kV/\cm$, and $\omega = 100 \ \kHz$. The green line corresponds to the projected reach of a detector optimized for detection of magnetic fields, such as the DM Radio experiment~\cite{Silva-Feaver:2016qhh}. The reach of dedicated LC resonators optimized for detecting electric fields is also shown. The lines labelled ``$E$-field (I-III)" correspond to various deflector/shield volumes, LC circuit temperatures, and quality factors as indicated in the legend. Also shown are the direct detection sensitivities of 1-year exposures for the near-term planned experiments SENSEI (100 g) (purple)~\cite{Crisler:2018gci,Abramoff:2019dfb} and SuperCDMS-G2 (1 kg) (yellow)~\cite{Battaglieri:2017aum}, assuming zero background.  Longer-term R\&D on direct detection concepts with meV-scale energy thresholds (such as detectors using superconductors~\cite{Hochberg:2015fth}, Dirac materials~\cite{Hochberg:2017wce}, or polar crystals~\cite{Knapen:2017ekk,Griffin:2018bjn} as targets) could extend direct detection sensitivity to keV-scale DM masses. Along the solid blue line, the millicharge abundance from freeze-in production in the early universe is in agreement with the observed dark matter energy density.}
\label{fig:reach}
\vspace{-0.5cm}
\end{figure}
\begin{figure}[t!]
\includegraphics[width = 0.5\textwidth]{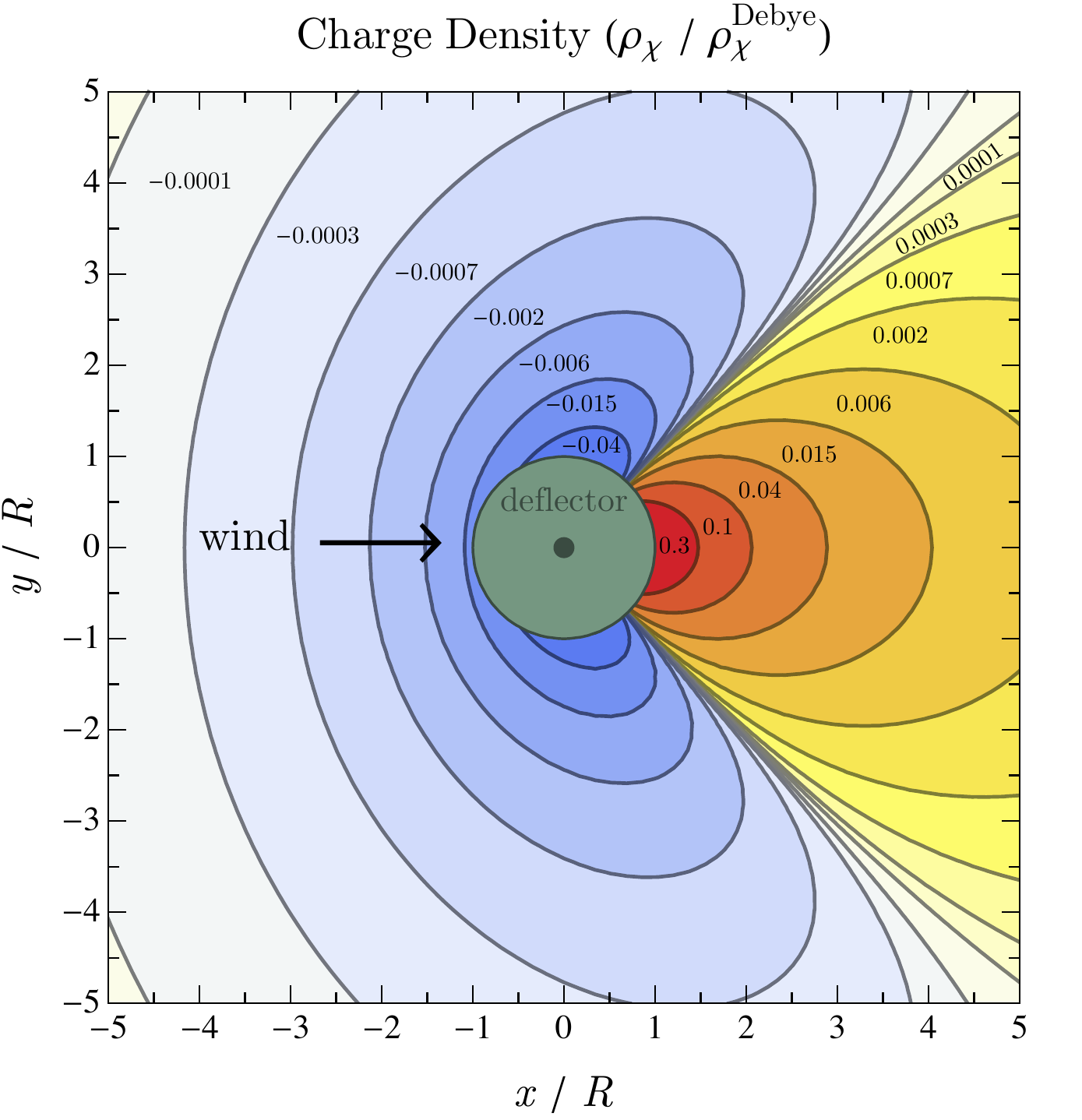}
\vspace{-0.7cm}
\caption{The time-independent amplitude of the dark matter charge density, $\rho_\x$, in the $x-y$ plane, induced by a deflector consisting of a point charge inside a spherical shield of radius $R$. The direction of the dark matter wind is along the positive $x$ direction, as specified by the black arrow. The red/blue contours correspond to positive/negative values of $\rho_\x$ when normalized by the dimensionful quantity $\rho_\x^\text{Debye}$ (see Eq.~(\ref{eq:debye2})). See the Supplementary Material for a more detailed discussion.
}
\label{fig:chargecontour}
\vspace{-0.5cm}
\end{figure}

Sensitivity to this range of mediator masses is well matched to models of sub-MeV DM production in the early universe. The primary benchmark model for production of sub-MeV DM is the ``freeze-in''~\cite{Hall:2009bx} of a DM abundance from the annihilations of thermal electrons~\cite{essig:2011nj,Chu:2011be,Knapen:2017xzo} (and a related reaction, plasmon decay~\cite{Dvorkin:2019zdi}).  
These reactions generate a DM abundance consistent with observations for couplings of size   
\be
\label{eq:freezein}
\qeff \sim \frac{1}{\alpha_\text{em}}  \left( \frac{m_e \, T_\text{eq}}{m_\x \, m_\text{Pl}} \right)^{1/2} \sim 10^{-10} \times\left( \frac{m_\x}{\keV} \right)^{-1/2}
\, ,
\ee
where $m_e$ is the electron mass, $m_\text{Pl}$ is the Planck mass, and $T_\text{eq} \simeq 0.8 \ \eV$ is the temperature at matter-radiation equality~\cite{Dvorkin:2019zdi}.  In order to remain consistent with other constraints, realizing this scenario for sub-MeV dark photon mediators requires $\mAp \lesssim 10^{-9} \ \eV$~\cite{Knapen:2017xzo,Dvorkin:2019zdi}. Therefore, viable freeze-in models for sub-MeV DM lie firmly in the millicharge-like regime for the class of experiments we consider.\footnote{It has been argued that millicharged DM may be evacuated from the galactic disk by supernova shocks~\cite{Chuzhoy:2008zy, McDermott:2010pa} (but see, e.g., Ref.~\cite{Dunsky:2018mqs} for claims to the contrary).  This effect is irrelevant for millicharge-like DM with $\mAp \gtrsim (100 \text{ pc})^{-1} \sim 10^{-25} \ \eV$, but 
might prevent all terrestrial experiments, including our concept, from detecting truly millicharged DM. For an investigation of other effects involving interactions between millicharged DM and galactic magnetic fields, see, e.g., Ref.~\cite{Stebbins:2019xjr}.}

\emph{Overview of Direct Deflection.}
A schematic illustration of the experimental setup is shown in Fig.~\ref{fig:setup}. A charge-symmetric, spatially uniform DM population passes through a shielded region of radius $R$, with an electric field\footnote{Since virialized DM is non-relativistic, millicharged DM is more efficiently deflected by electric (rather than magnetic) fields.} oscillating at angular frequency $\omega$. We refer to this region as the ``deflector."  The velocity distribution of DM in the earth's frame is expected to be approximately Maxwellian, shifted by a ``wind'' velocity, $v_\text{wind}$, from the sun's motion in the Milky Way, with velocity dispersion $v_0 \sim v_\text{wind} \sim \order{100} \ \km / \s$~\cite{Freese:2012xd}.  

As millicharged DM passes through the deflector, it is subject to an electric force that separates positively and negatively charged particles. This creates a propagating wave train of alternating millicharge ($\rho_\x$) and millicurrent ($j_\x$) densities of length $\sim 2 \pi v_\text{wind} / \omega$, which diffuse outwards due to dispersion in the DM velocity distribution. DM particles easily penetrate electromagnetic shielding, inducing small electromagnetic fields within the shielded detection region of Fig.~\ref{fig:setup}.  These fields have known oscillation frequency and phase, and can be measured using an electric field pickup antenna coupled to a resonant LC circuit and SQUID amplifier.  Relative to the DM wind (due to the sun's galactocentric velocity), the apparatus rotates once per sidereal day; we have illustrated the geometry where the signal is maximized in Fig.~\ref{fig:setup}. 
 
\emph{Inducing Dark Matter Waves.}
Oscillating the sign of the deflector field allows for resonant read-out, but the oscillations should be slow enough that DM particles traverse the deflector within one period, i.e., 
\be
\label{eq:freq}
\omega \lesssim \pi \, v_\chi / R \sim \MHz \times \left( R / \text{meter} \right)^{-1} 
\, ,
\ee
where we take the characteristic DM velocity to be $v_\x \sim v_{0 , \text{wind}} \sim 10^{-3} c$. When Eq.~(\ref{eq:freq}) is satisfied, we can treat the fields as quasi-static.  Moreover, astrophysical bounds on DM self-interactions constrain the Debye length of the DM plasma to be  $\gtrsim 100 \text{ meters}$ for $m_\x \gtrsim \eV$~\cite{Lin:2019uvt}, guaranteeing that collisional/backreaction effects in the DM plasma can be neglected when modeling the apparatus of Fig.~\ref{fig:setup}. 

In the Supplementary Material, we derive millicharge and millicurrent densities for general DM velocity distributions and deflector charge distributions.  The DM millicharge density can be simply expressed in terms of general (multipole and trace) moments of the deflector charge distribution.  
For any deflector surrounded by a grounded shield, all multipole moments vanish (as required to obtain vanishing electric fields outside the shield).  The induced DM charge density far outside the shield ($|\xv| \gg R$) is therefore dominated by the leading trace moment of the deflector-plus-shield charge distribution, which is its charge-radius-squared ($\mathcal{R}_\defl^2$), i.e.,
\begin{align}
\label{eq:simplerExact}
\rho_\x (\xv) &\simeq - \frac{(e \qeff)^2 \, \rhodm \, \mathcal{R}_\defl^2 }{6 m_\x^2}  
\int dv  ~ \lap^2 \frac{f(v \, \hat{\xv})}{|\xv|} 
~,
\end{align}
where $\rhodm \simeq 0.4 \ \GeV/\cm^3$ is the local DM energy density, $f(\vv)$ is the DM velocity distribution in the lab frame (which we take to be a shifted-Maxwellian), and we have taken the deflector to be centered at the origin. The velocity integral in Eq.~(\ref{eq:simplerExact}) scales with distance as $1/|\xv|^3$.

For a given enclosed deflector charge, $q_\defl$, and spherical shield of radius $R$, a centered point charge maximizes the charge-radius of the shielded deflection region and therefore represents the optimal deflector.  The DM charge density induced by this geometry is shown in Fig.~\ref{fig:chargecontour}.  Similar DM charge densities are obtained for any shielded deflector with non-vanishing charge-radius, such as a spherical or cylindrical capacitor of size comparable to $R$, or a dipole or parallel plate capacitor that is substantially displaced from the center of the surrounding shield. In contrast, the DM charge overdensities resulting from a parallel plate capacitor centered in the shield (which has vanishing charge-radius) are considerably smaller in the $|\xv| \gg R$ limit.

The distance-scaling and angular profile of the result in Eq.~(\ref{eq:simplerExact}) and Fig.~\ref{fig:chargecontour} can be understood as follows: 
The millicharge distribution induced by an \emph{unshielded} point charge deflector is approximately given by Debye screening,
\be
\label{eq:debye0}
\rho_\x^\text{Debye}(\xv) \sim - \frac{(e \qeff)^2 \, \rhodm \, \phi_\defl(\xv)}{m_\x^2 \, v_0^2} 
~,
\ee
with $\phi_\defl (\xv) \sim e q_\defl / |\xv|$.  The charge-radius-squared that induces the leading $\rho_\x$  for a \emph{shielded} deflector (see Eq.~(\ref{eq:simplerExact})) is a second moment, and so the resulting form of $\rho_\x$ must be further suppressed by $R^2/|\xv|^2$. 

Since the charge-radius doesn't contribute to the electric potential outside the deflector, its effect must vanish in the no-wind limit where the Debye screening result of Eq.~(\ref{eq:debye0}) applies exactly.  Indeed, we can think of Debye screening as gradually building up a millicharge distribution inside the shield, which is then both ``dragged'' downwind at a velocity $v_\text{wind}$ and dispersed outwards with a characteristic velocity $v_0$.  Therefore, a diluted millicharge density with the same sign as Eq.~(\ref{eq:debye0}) arises in a cone of angular size $v_0/v_\text{wind}$ about the downwind direction (the $R^2/|\xv|^2$ suppression relative to Debye screening can be physically understood as resulting from the dilution of charge starting from the transverse size $R$ of the source to the transverse size $(v_0/v_\text{wind}) |\xv|$ of the cone), with an opposite-charged density outside of this cone due to charge conservation.  For $v_\text{wind} \gtrsim v_0$, we find in the downwind region
\be
\label{eq:estimate1}
\rho_\x (\xv) \sim\rho_\x^\text{Debye}(R) \left( \frac{v_\text{wind}}{v_0}\right)^2 \left(\frac{R}{|\xv|} \right)^3 
~.
\ee
Since these DM charge densities travel in the lab frame at a speed $\sim v_\text{wind}$, the corresponding current densities are roughly,
\be
\label{eq:jestimate}
j_\x (\xv) \sim \rho_\x (\xv) \, v_\text{wind}
~.
\ee
For the optimal geometry of Fig.~\ref{fig:setup}, in which the DM wind is aligned with the detection region, the parametric expressions in Eqs.~(\ref{eq:estimate1}) and (\ref{eq:jestimate}) agree within $\order{1}$ factors with the detailed calculations presented in the Supplementary Material. 

\emph{Detecting Dark Matter Waves.} 
As shown in the setup of Fig.~\ref{fig:setup}, to detect the DM millicharge or millicurrent densities, a detector is placed downstream of the deflector.  The detector should be surrounded by its own electromagnetic shield, in order to reduce noise from external non-DM sources. We assume that the characteristic length scale of this shielded detection region is comparable to that of the deflector and is therefore also in the quasi-static limit. For concreteness, we take the detector shield to be a sphere of radius $R$.

The charge and current densities oscillate inside the detector shield at the same frequency as the deflector, 
\be
\label{eq:rhoandj}
\rho_\x (t) \simeq \rho_\x \, e^{i \omega t} ~,~\jx (t) \simeq \jx \, e^{i \omega t}
~.
\ee
Oscillating charge and current densities inside a conducting shield generate visible electric ($E_\x$) and magnetic ($B_\x$) fields that oscillate at the same frequency. In the quasi-static limit ($\omega \ll 1/R$), the dominant effect of the charge/current density is a small electric/magnetic field. We numerically calculate these fields inside the detector shield using the results for $\rho_\x$ and $\jv_\x$ presented in the Supplementary Material. Parametrically, we expect $E_\x \sim \rho_\x R$ and $B_\x \sim j_\x R$. Eq.~(\ref{eq:jestimate}) therefore implies that the electric field sourced by the DM charge density is \emph{velocity-enhanced} compared to the magnetic field sourced by the current density ($E_\x \sim v_\x^{-1} \, B_\x$), as expected for non-relativistic charge carriers. 

An excellent detector for these DM-sourced electromagnetic fields is a well-shielded resonant LC circuit and antenna, inductively coupled to a SQUID.  LC circuits can operate at frequencies much smaller than the inverse geometric size of their corresponding circuit components ($\omega_\text{LC} = 1 / \sqrt{LC} \ll 1 / R$), in contrast to, e.g., superconducting RF cavities.\footnote{It is interesting to note that recent work has investigated the possibility that a subcomponent of the millicharged DM population could be accelerated to semi-relativistic speeds by supernova remnants~\cite{Chuzhoy:2008zy,Hu:2016xas,Dunsky:2018mqs}. Depending on the abundance of such ``dark cosmic rays," a setup similar to that shown in Fig.~\ref{fig:setup} involving a pair of resonant SRF cavities could have greatly enhanced sensitivity due to their extremely large field gradients and quality factors.}
If the LC circuit is resonantly matched to the deflector frequency ($\omega_\text{LC} \simeq \omega$), then in the presence of the DM charge and current densities, the LC circuit responds by ringing up the small oscillating fields over many cycles, quantified by its large quality factor, $Q \gg 1$. Such technology has been extensively developed and tested by AURIGA~\cite{Cerdonio:1997hz} to detect gravity waves and will be implemented by DM Radio~\cite{Silva-Feaver:2016qhh} to search for ultralight coherent DM fields. 

An LC resonator with a large inductor, such as that proposed for the DM Radio experiment, couples mainly to the oscillating magnetic fields in the detector sourced by DM currents~\cite{Silva-Feaver:2016qhh}.  Detecting the parametrically larger electric field signals from DM charges motivates an LC resonator incorporating a large capacitor or receiving antenna.\footnote{This is in contrast to the case of axion DM, where electric signals have been considered~\cite{McAllister:2018ndu} but are suppressed relative to the magnetic one~\cite{Ouellet:2018nfr,Beutter:2018xfx}.  The difference in scaling arises because in our setup, electromagnetic sources (millicharged DM particles) are present inside the shielded detection region.}   In this work, we consider both inductively and capacitatively coupled detectors. We assume that stray electromagnetic noise is sufficiently well-shielded that the experiment is limited by thermal (Johnson-Nyquist) noise in the detector, with a power spectral density equal to $4 R_\text{LC} \, T_\text{LC}$, where $R_\text{LC}$ and $T_\text{LC}$ are the resistance and temperature of the LC resonator, respectively~\cite{Nyquist:1928zz}.  The shielding requirements are similar to  those for ``light-shining-through-a-wall'' cavity experiments~\cite{Jaeckel:2007ch,Betz:2013dza,Graham:2014sha} or DM Radio~\cite{Silva-Feaver:2016qhh}.

The signal-to-noise ratio (SNR) is given by the corresponding ratio of power spectral densities. For an experiment limited by thermal noise and assuming an integration time $\tint$ that is much shorter than the phase-coherence time of the deflector, the SNR is approximately~\cite{Budker:2013hfa,Graham:2014sha}
\be
\label{eq:SNR}
\text{SNR} \simeq \frac{\omega \, Q \, t_\text{int}}{4 \, T_\text{LC}} \, \int_\text{det} d^3 \xv ~ (E_\x^2 \text{ or } B_\x^2)
\propto \left(\frac{\qeff}{m_\x}\right)^4
\, .
\ee
The integral of $E_\x^2$ or $B_\x^2$ is over an effective detector volume (the physical volume of a resonant capacitor or inductor, or the effective volume of a receiving antenna).  In either case, these volumes are bounded by the total volume of the shielded detector cavity, $V_\text{sh}$. A receiving antenna pickup allows for the physical volume of the LC circuit to be significantly smaller than $V_\text{sh}$, giving a considerable advantage in terms of cooling power.

\emph{Expected Reach.} 
The projected reach of our setup scales as\footnote{For deflector and detector shields of different volumes, the reach in Eq.~(\ref{eq:reach}) scales as $V_\text{def}^{-1/6} \, V_\text{det}^{-5/12}$.}
\be
\label{eq:reach}
\qeff / m_\x \propto V_\text{sh}^{-7/12} \, \langle E_\defl \rangle^{-1/2} \, T_\text{LC}^{1/4} \, ( \omega \, t_\text{int} \, Q)^{-1/4}
~,
\ee
which is mainly sensitive to the deflector/shield volume, $V_\text{sh}$, and the spatially-averaged strength of the deflecting field, $\langle E_\defl \rangle$. The estimated sensitivities for different experimental configurations are shown in Fig.~\ref{fig:reach}, compared to existing constraints (shaded gray)~\cite{Essig:2012yx,Essig:2017kqs,Vogel:2013raa,Chang:2018rso,Kovetz:2018zan,Williams:1971ms}. In all cases, we assume a year of integration time and a deflector operating with a spatially-averaged field-strength of $\langle E_\defl \rangle = 10 \ \kV/\cm$ at a frequency of $\omega = 100 \ \kHz$. 
Thermal voltage fluctuations due to Johnson-Nyquist noise are randomly distributed according to a Gaussian centered at zero. Since this implies that power fluctuations follow a rescaled chi-squared distribution, we estimate the thermal noise limited reach of an LC resonator by requiring $\text{SNR} \gtrsim 2.7$, corresponding to a 90\% confidence limit. 

The green line in Fig.~\ref{fig:reach} shows the projected reach of the planned DM Radio experiment, which is optimized to detect small magnetic fields and is expected to consist of a $\text{meter}^3$ detector with $T_\text{LC} = 10 \ \mK$ and $Q=10^6$, assuming that it is modified to include a $\text{meter}^3$ upstream deflector. 
The construction of thermal-noise-limited resonant LC circuits operating at kHz frequencies with quality factors of $Q \simeq 10^6$ has already been firmly established by the existing gravity-wave experiment, AURIGA~\cite{Baggio:2005xp,Bonaldi:1998gcg,Bonaldi:1999mvu}. Therefore, we also consider a future dedicated experimental configuration optimized for measuring the electric field signal, as shown by the lines labelled ``$E$-field (I-III)" in Fig.~\ref{fig:reach}. These correspond to deflector/shield volumes, LC circuit temperatures, and quality factors ranging from $V_\text{sh} = (5-10) \ \text{meter}^3$, $T_\text{LC} = 100 \ \mK - 4 \ \text{K}$, and $Q = 10^4 - 10^7$, as noted in the legend of Fig.~\ref{fig:reach}. In particular, the ``$E$-field (I)" configuration demonstrates that even a conservative setup could attain impressive sensitivity to cosmologically motivated parameter space. Compared to AURIGA and DM Radio, our setup does not need to scan or operate at frequencies below a $\kHz$, which often necessitates the use of lossy dielectrics. Hence, quality factors greater than $10^6$ might be more easily attainable.  

In Fig.~\ref{fig:reach}, we have assumed that the separation distance between the deflector and detector is small compared to their overall size (a geometrical penalty as in Eq.~(\ref{eq:estimate1}) still exists because of the finite sizes of the deflector and detector). We note that spherical geometries, such as that used in Ref.~\cite{Williams:1971ms}, can potentially overcome this minor penalty. We have also time-averaged the power in Eq.~(\ref{eq:SNR}) over a sidereal day, assuming that the axis joining the deflector and detector regions of Fig.~\ref{fig:setup} is aligned with the direction of the DM wind at time $t = 0$ (see the discussion below and in, e.g., Ref.~\cite{Griffin:2018bjn}). 

In order to maximally ring up the signal electromagnetic field in the LC resonator, it is important to keep the deflector and the response of the LC circuit approximately in phase.  The phase drift of the driven deflector field can be minimized if it is phase-locked to a precise external clock. The phase control does not have to be especially precise --- phase differences contained within a range $\pm \delta \phi\ll 1 $ only degrade the signal by $1 - \order{\delta\phi^2}$ --- but should prevent drift of the deflector phase by a full period or more over the data-taking time.  Ensuring this level of phase stability over one year of running at $100 \ \kHz$ calls for a clock precision of $ \sim 10^{-12}$, which is readily achieved using commonly available (remotely linked) NIST reference clocks~\cite{Heavner_2014}.  
Similarly, Poisson fluctuations in the local DM density lead to an irreducible stochastic modulation of signal power, even for a well phase-locked deflector.   For DM masses below a GeV, the relative variations are at the level of $\lesssim 10^{-3}$ and so most of the signal power is not spread out in frequency by these fluctuations.  Neither of the above effects leads to long-term loss of coherence.

\emph{Daily Modulation.} 
The directionality of the DM wind provides an additional handle to discriminate a DM signal from unforeseen systematics or noise at the deflector frequency: the earth's rotation leads to a daily modulation of the angle between the detector-deflector axis and the DM wind.  The strength of the signal is maximized (minimized) when the detector-deflector axis is aligned (anti-aligned) with the wind. This effect introduces a strong directionality as well as a large ($\gg \order{1}$) fractional daily modulation to the signal as the variation of the DM wind follows the rotation of the earth, both of which can help to identify a true signal (see the left panel of Fig.~\ref{fig:angleandcurrent}). In a simple Fourier analysis, most of the observed signal power will be shifted from the deflector frequency, $\omega$, to $\omega \pm \omega_\oplus$, where $\omega_\oplus$ is the frequency of a sidereal day --- in contrast to deflector-induced backgrounds which are expected only at frequency $\omega$.  Relatedly, while deflector leakage at frequency $\omega$ would be reduced by convolving the signal with a template accounting for the daily rotation of the detector, a DM signal would be \emph{enhanced} by this procedure. These frequency differences are well within the detector bandwidth for realistic quality factors, but nonetheless clearly distinguishable over integration times longer than a few days. The precise form of this daily modulation and its dependence on the detector orientation are sensitive probes of the local DM velocity distribution, $f(\vv)$.  We have taken $f(\vv)$ to be a shifted-Maxwellian for definiteness, but additional sources of anisotropy such as debris flow~\cite{Necib:2018iwb} would change the detailed daily-modulation pattern of the signal.

\emph{Comparison to Direct Detection.} 
In Fig.~\ref{fig:reach}, we show as dotted lines the projected sensitivities of two representative direct detection searches for energy deposition from DM scattering: a near-term $2 e^-$, 100~g-year run of the SENSEI experiment (purple)~\cite{Crisler:2018gci,Abramoff:2019dfb} and a kg-year exposure for SuperCDMS-G2 (yellow)~\cite{Battaglieri:2017aum}.   These near-term experiments have MeV-scale DM mass thresholds (arising from eV-scale electron recoil energy thresholds), and are highly complementary to the low-mass sensitivity of the  direct \emph{deflection} setup proposed here.  R\&D efforts towards developing technologies  capable of detecting keV-scale DM include proposals to use sapphire~\cite{Knapen:2017ekk,Griffin:2018bjn}, Dirac materials~\cite{Hochberg:2017wce}, or superconductors~\cite{Hochberg:2015fth} as targets. These would provide an independent test of the cosmologically motivated freeze-in parameter space (blue line) in the $\keV-\MeV$ mass-range, as well as other DM models with shorter-range forces. 

A unique qualitative feature of the direct deflection technique discussed here is that the sensitivity improves at smaller masses,  
corresponding to larger DM number densities and smaller momentum carried by individual DM particles.
Probing the collective effects of the large number density of DM particles, instead of relying on the energy deposition from a single DM scattering event, allows for LC resonators coupled to electromagnetic deflectors to probe parameter space for DM masses well below the keV-scale. We note that cosmological and astrophysical constraints have not yet been extensively investigated for such small masses; we leave the detailed investigation of these effects to future work. 

In Fig.~\ref{fig:reach}, we have shown the sensitivity of another type of experiment. Ref.~\cite{Mitra:2006ds} noted that in the presence of a millicharged DM plasma, the test of Coulomb's law in Ref.~\cite{Williams:1971ms} would detect a positive signal in the form of an oscillating voltage gradient across a spherical capacitor. The effect is nearly identical to the one discussed in this work, but a rough estimate of the bound (dashed gray in Fig.~\ref{fig:reach}) shows that it is currently superseded by cosmological and astrophysical constraints.  

We also note that our classical calculations are not valid to arbitrarily small DM masses. For $m_\x \lesssim 0.1 \text{ meV}$,
the de Broglie wavelength of an incoming DM particle is $(m_\x v_\x)^{-1} \gtrsim \order{\text{meter}}$, and hence comparable to the deflector size. In this regime (not shown in Fig.~\ref{fig:reach}), a quantum treatment of millicharge and millicurrent production is required.

\emph{Discussion.} 
We have proposed a new approach for the direct detection of sub-MeV DM that couples weakly to electromagnetism. This experimental technique applies technology that has been developed to search for ultralight coherent fields and has the potential to gain orders of magnitude in sensitivity to millicharged (and millicharge-like) DM.  

Our setup highlights the ability to detect DM in the Milky Way halo by inducing collective disturbances into the DM population through the manipulation of strong electromagnetic fields in the lab. Similar techniques can be more generally applied beyond what is considered in this work. Such ideas include, e.g., the application of specifically engineered field configurations to accelerate, focus, or trap DM near standard direct detection scattering targets or resonant detectors. We leave such considerations to future work.

Furthermore, generalizations of the proposed experimental setup with a similar ``deflection-detection" approach could be used to search for alternative types of DM-SM interactions. For example, oscillating spin-polarized samples (such as those planned for the ARIADNE axion experiment~\cite{Geraci:2017bmq}) could be used to deflect and detect particle DM that interacts through macroscopic spin-coupled forces. For these reasons, \emph{direct deflection} of DM constitutes a promising alternative avenue towards the discovery of sub-MeV DM using established experimental techniques.

{\emph{Acknowledgments.} ---} 
We thank Tony Beukers, Saptarshi Chaudhuri, Paolo Falferi, Peter Graham, Yoni Kahn, Tongyan Lin, Michael Peskin, Tor Raubenheimer, Jesse Thaler, 
Andrea Vinante, and Marco Zanetti for valuable discussions. The authors are supported by the U.S. Department of Energy under Contract No. DE-AC02-76SF00515.


\bibliography{MilliQ}

\begin{thebibliography}{10}%
\makeatletter
\providecommand \@ifxundefined [1]{%
 \ifx #1\undefined \expandafter \@firstoftwo
 \else \expandafter \@secondoftwo
\fi
}%
\providecommand \@ifnum [1]{%
 \ifnum #1\expandafter \@firstoftwo
 \else \expandafter \@secondoftwo
\fi
}%
\providecommand \enquote [1]{``#1''}%
\providecommand \bibnamefont  [1]{#1}%
\providecommand \bibfnamefont [1]{#1}%
\providecommand \citenamefont [1]{#1}%
\providecommand\href[0]{\@sanitize\@href}%
\providecommand\@href[1]{\endgroup\@@startlink{#1}\endgroup\@@href}%
\providecommand\@@href[1]{#1\@@endlink}%
\providecommand \@sanitize [0]{\begingroup\catcode`\&12\catcode`\#12\relax}%
\@ifxundefined \pdfoutput {\@firstoftwo}{%
 \@ifnum{\z@=\pdfoutput}{\@firstoftwo}{\@secondoftwo}%
}{%
 \providecommand\@@startlink[1]{\leavevmode\special{html:<a href="#1">}}%
 \providecommand\@@endlink[0]{\special{html:</a>}}%
}{%
 \providecommand\@@startlink[1]{%
  \leavevmode
  \pdfstartlink
   attr{/Border[0 0 1 ]/H/I/C[0 1 1]}%
   user{/Subtype/Link/A<</Type/Action/S/URI/URI(#1)>>}%
  \relax
 }%
 \providecommand\@@endlink[0]{\pdfendlink}%
}%
\providecommand \url  [0]{\begingroup\@sanitize \@url }%
\providecommand \@url [1]{\endgroup\@href {#1}{\urlprefix}}%
\providecommand \urlprefix [0]{URL }%
\providecommand \Eprint[0]{\href }%
\@ifxundefined \urlstyle {%
  \providecommand \doi [1]{doi:\discretionary{}{}{}#1}%
}{%
  \providecommand \doi [0]{doi:\discretionary{}{}{}\begingroup
  \urlstyle{rm}\Url }%
}%
\providecommand \doibase [0]{http://dx.doi.org/}%
\providecommand \Doi[1]{\href{\doibase#1}}%
\providecommand \bibAnnote [3]{%
  \BibitemShut{#1}%
  \begin{quotation}\noindent
    \textsc{Key:}\ #2\\\textsc{Annotation:}\ #3%
  \end{quotation}%
}%
\providecommand \bibAnnoteFile [2]{%
  \IfFileExists{#2}{\bibAnnote {#1} {#2} {\input{#2}}}{}%
}%
\providecommand \typeout [0]{\immediate \write \m@ne }%
\providecommand \selectlanguage [0]{\@gobble}%
\providecommand \bibinfo [0]{\@secondoftwo}%
\providecommand \bibfield [0]{\@secondoftwo}%
\providecommand \translation [1]{[#1]}%
\providecommand \BibitemOpen[0]{}%
\providecommand \bibitemStop [0]{}%
\providecommand \bibitemNoStop [0]{.\EOS\space}%
\providecommand \EOS [0]{\spacefactor3000\relax}%
\providecommand \BibitemShut [1]{\csname bibitem#1\endcsname}%
\bibitem{Battaglieri:2017aum}%
  \BibitemOpen
  \bibfield{author}{%
  \bibinfo {author} {\bibfnamefont{M.}~\bibnamefont{Battaglieri}}
  \emph{et~al.},\ }%
  in\ \emph{\bibinfo {booktitle} {{U.S. Cosmic Visions: New Ideas in Dark
  Matter College Park, MD, USA, March 23-25, 2017}}}\ (\bibinfo {year} {2017})\
  \Eprint{http://arxiv.org/abs/1707.04591}{arXiv:1707.04591 [hep-ph]},\
  \url{http://lss.fnal.gov/archive/2017/conf/fermilab-conf-17-282-ae-ppd-t.pdf%
}%
  \bibAnnoteFile{NoStop}{Battaglieri:2017aum}%
\bibitem{Hall:2009bx}%
  \BibitemOpen
  \bibfield{author}{%
  \bibinfo {author} {\bibfnamefont{L.~J.}\ \bibnamefont{Hall}}, \bibinfo
  {author} {\bibfnamefont{K.}~\bibnamefont{Jedamzik}}, \bibinfo {author}
  {\bibfnamefont{J.}~\bibnamefont{March-Russell}},\ and\ \bibinfo {author}
  {\bibfnamefont{S.~M.}\ \bibnamefont{West}},\ }%
  \bibfield{journal}{%
  \Doi{10.1007/JHEP03(2010)080}{\bibinfo {journal} {JHEP}}\ }%
  \textbf{\bibinfo {volume} {03}},\ \bibinfo {pages} {080} (\bibinfo {year}
  {2010}),\ \Eprint{http://arxiv.org/abs/0911.1120}{arXiv:0911.1120 [hep-ph]}%
  \bibAnnoteFile{NoStop}{Hall:2009bx}%
\bibitem{Jaeckel:2007ch}%
  \BibitemOpen
  \bibfield{author}{%
  \bibinfo {author} {\bibfnamefont{J.}~\bibnamefont{Jaeckel}}\ and\ \bibinfo
  {author} {\bibfnamefont{A.}~\bibnamefont{Ringwald}},\ }%
  \bibfield{journal}{%
  \Doi{10.1016/j.physletb.2007.11.071}{\bibinfo {journal} {Phys. Lett.}}\ }%
  \textbf{\bibinfo {volume} {B659}},\ \bibinfo {pages} {509} (\bibinfo {year}
  {2008}),\ \Eprint{http://arxiv.org/abs/0707.2063}{arXiv:0707.2063 [hep-ph]}%
  \bibAnnoteFile{NoStop}{Jaeckel:2007ch}%
\bibitem{Betz:2013dza}%
  \BibitemOpen
  \bibfield{author}{%
  \bibinfo {author} {\bibfnamefont{M.}~\bibnamefont{Betz}}, \bibinfo {author}
  {\bibfnamefont{F.}~\bibnamefont{Caspers}}, \bibinfo {author}
  {\bibfnamefont{M.}~\bibnamefont{Gasior}}, \bibinfo {author}
  {\bibfnamefont{M.}~\bibnamefont{Thumm}},\ and\ \bibinfo {author}
  {\bibfnamefont{S.~W.}\ \bibnamefont{Rieger}},\ }%
  \bibfield{journal}{%
  \Doi{10.1103/PhysRevD.88.075014}{\bibinfo {journal} {Phys. Rev.}}\ }%
  \textbf{\bibinfo {volume} {D88}},\ \bibinfo {pages} {075014} (\bibinfo {year}
  {2013}),\ \Eprint{http://arxiv.org/abs/1310.8098}{arXiv:1310.8098
  [physics.ins-det]}%
  \bibAnnoteFile{NoStop}{Betz:2013dza}%
\bibitem{Graham:2014sha}%
  \BibitemOpen
  \bibfield{author}{%
  \bibinfo {author} {\bibfnamefont{P.~W.}\ \bibnamefont{Graham}}, \bibinfo
  {author} {\bibfnamefont{J.}~\bibnamefont{Mardon}}, \bibinfo {author}
  {\bibfnamefont{S.}~\bibnamefont{Rajendran}},\ and\ \bibinfo {author}
  {\bibfnamefont{Y.}~\bibnamefont{Zhao}},\ }%
  \bibfield{journal}{%
  \Doi{10.1103/PhysRevD.90.075017}{\bibinfo {journal} {Phys. Rev.}}\ }%
  \textbf{\bibinfo {volume} {D90}},\ \bibinfo {pages} {075017} (\bibinfo {year}
  {2014}),\ \Eprint{http://arxiv.org/abs/1407.4806}{arXiv:1407.4806 [hep-ph]}%
  \bibAnnoteFile{NoStop}{Graham:2014sha}%
\bibitem{Holdom:1985ag}%
  \BibitemOpen
  \bibfield{author}{%
  \bibinfo {author} {\bibfnamefont{B.}~\bibnamefont{Holdom}},\ }%
  \bibfield{journal}{%
  \Doi{10.1016/0370-2693(86)91377-8}{\bibinfo {journal} {Phys. Lett.}}\ }%
  \textbf{\bibinfo {volume} {166B}},\ \bibinfo {pages} {196} (\bibinfo {year}
  {1986})%
  \bibAnnoteFile{NoStop}{Holdom:1985ag}%
\bibitem{Silva-Feaver:2016qhh}%
  \BibitemOpen
  \bibfield{author}{%
  \bibinfo {author} {\bibfnamefont{M.}~\bibnamefont{Silva-Feaver}}
  \emph{et~al.},\ }%
  \bibfield{booktitle}{%
  \emph{\bibinfo {booktitle} {{Proceedings, Applied Superconductivity
  Conference (ASC 2016): Denver, Colorado, September 4-9, 2016}}},\ }%
  \bibfield{journal}{%
  \Doi{10.1109/TASC.2016.2631425}{\bibinfo {journal} {IEEE Trans. Appl.
  Supercond.}}\ }%
  \textbf{\bibinfo {volume} {27}},\ \bibinfo {pages} {1400204} (\bibinfo {year}
  {2017}),\ \Eprint{http://arxiv.org/abs/1610.09344}{arXiv:1610.09344
  [astro-ph.IM]}%
  \bibAnnoteFile{NoStop}{Silva-Feaver:2016qhh}%
\bibitem{Crisler:2018gci}%
  \BibitemOpen
  \bibfield{author}{%
  \bibinfo {author} {\bibfnamefont{M.}~\bibnamefont{Crisler}}, \bibinfo
  {author} {\bibfnamefont{R.}~\bibnamefont{Essig}}, \bibinfo {author}
  {\bibfnamefont{J.}~\bibnamefont{Estrada}}, \bibinfo {author}
  {\bibfnamefont{G.}~\bibnamefont{Fernandez}}, \bibinfo {author}
  {\bibfnamefont{J.}~\bibnamefont{Tiffenberg}}, \bibinfo {author}
  {\bibfnamefont{M.}~\bibnamefont{Sofo~haro}}, \bibinfo {author}
  {\bibfnamefont{T.}~\bibnamefont{Volansky}},\ and\ \bibinfo {author}
  {\bibfnamefont{T.-T.}\ \bibnamefont{Yu}} (\bibinfo {collaboration}
  {SENSEI}),\ }%
  \bibfield{journal}{%
  \Doi{10.1103/PhysRevLett.121.061803}{\bibinfo {journal} {Phys. Rev. Lett.}}\
  }%
  \textbf{\bibinfo {volume} {121}},\ \bibinfo {pages} {061803} (\bibinfo {year}
  {2018}),\ \Eprint{http://arxiv.org/abs/1804.00088}{arXiv:1804.00088
  [hep-ex]}%
  \bibAnnoteFile{NoStop}{Crisler:2018gci}%
\bibitem{Abramoff:2019dfb}%
  \BibitemOpen
  \bibfield{author}{%
  \bibinfo {author} {\bibfnamefont{O.}~\bibnamefont{Abramoff}} \emph{et~al.}
  (\bibinfo {collaboration} {SENSEI})}%
   (\bibinfo {year} {2019}),\
  \Eprint{http://arxiv.org/abs/1901.10478}{arXiv:1901.10478 [hep-ex]}%
  \bibAnnoteFile{NoStop}{Abramoff:2019dfb}%
\bibitem{Hochberg:2015fth}%
  \BibitemOpen
  \bibfield{author}{%
  \bibinfo {author} {\bibfnamefont{Y.}~\bibnamefont{Hochberg}}, \bibinfo
  {author} {\bibfnamefont{M.}~\bibnamefont{Pyle}}, \bibinfo {author}
  {\bibfnamefont{Y.}~\bibnamefont{Zhao}},\ and\ \bibinfo {author}
  {\bibfnamefont{K.~M.}\ \bibnamefont{Zurek}},\ }%
  \bibfield{journal}{%
  \Doi{10.1007/JHEP08(2016)057}{\bibinfo {journal} {JHEP}}\ }%
  \textbf{\bibinfo {volume} {08}},\ \bibinfo {pages} {057} (\bibinfo {year}
  {2016}),\ \Eprint{http://arxiv.org/abs/1512.04533}{arXiv:1512.04533
  [hep-ph]}%
  \bibAnnoteFile{NoStop}{Hochberg:2015fth}%
\bibitem{Hochberg:2017wce}%
  \BibitemOpen
  \bibfield{author}{%
  \bibinfo {author} {\bibfnamefont{Y.}~\bibnamefont{Hochberg}}, \bibinfo
  {author} {\bibfnamefont{Y.}~\bibnamefont{Kahn}}, \bibinfo {author}
  {\bibfnamefont{M.}~\bibnamefont{Lisanti}}, \bibinfo {author}
  {\bibfnamefont{K.~M.}\ \bibnamefont{Zurek}}, \bibinfo {author}
  {\bibfnamefont{A.~G.}\ \bibnamefont{Grushin}}, \bibinfo {author}
  {\bibfnamefont{R.}~\bibnamefont{Ilan}}, \bibinfo {author}
  {\bibfnamefont{S.~M.}\ \bibnamefont{Griffin}}, \bibinfo {author}
  {\bibfnamefont{Z.-F.}\ \bibnamefont{Liu}}, \bibinfo {author}
  {\bibfnamefont{S.~F.}\ \bibnamefont{Weber}},\ and\ \bibinfo {author}
  {\bibfnamefont{J.~B.}\ \bibnamefont{Neaton}},\ }%
  \bibfield{journal}{%
  \Doi{10.1103/PhysRevD.97.015004}{\bibinfo {journal} {Phys. Rev.}}\ }%
  \textbf{\bibinfo {volume} {D97}},\ \bibinfo {pages} {015004} (\bibinfo {year}
  {2018}),\ \Eprint{http://arxiv.org/abs/1708.08929}{arXiv:1708.08929
  [hep-ph]}%
  \bibAnnoteFile{NoStop}{Hochberg:2017wce}%
\bibitem{Knapen:2017ekk}%
  \BibitemOpen
  \bibfield{author}{%
  \bibinfo {author} {\bibfnamefont{S.}~\bibnamefont{Knapen}}, \bibinfo {author}
  {\bibfnamefont{T.}~\bibnamefont{Lin}}, \bibinfo {author}
  {\bibfnamefont{M.}~\bibnamefont{Pyle}},\ and\ \bibinfo {author}
  {\bibfnamefont{K.~M.}\ \bibnamefont{Zurek}},\ }%
  \bibfield{journal}{%
  \Doi{10.1016/j.physletb.2018.08.064}{\bibinfo {journal} {Phys. Lett.}}\ }%
  \textbf{\bibinfo {volume} {B785}},\ \bibinfo {pages} {386} (\bibinfo {year}
  {2018}),\ \Eprint{http://arxiv.org/abs/1712.06598}{arXiv:1712.06598
  [hep-ph]}%
  \bibAnnoteFile{NoStop}{Knapen:2017ekk}%
\bibitem{Griffin:2018bjn}%
  \BibitemOpen
  \bibfield{author}{%
  \bibinfo {author} {\bibfnamefont{S.}~\bibnamefont{Griffin}}, \bibinfo
  {author} {\bibfnamefont{S.}~\bibnamefont{Knapen}}, \bibinfo {author}
  {\bibfnamefont{T.}~\bibnamefont{Lin}},\ and\ \bibinfo {author}
  {\bibfnamefont{K.~M.}\ \bibnamefont{Zurek}},\ }%
  \bibfield{journal}{%
  \Doi{10.1103/PhysRevD.98.115034}{\bibinfo {journal} {Phys. Rev.}}\ }%
  \textbf{\bibinfo {volume} {D98}},\ \bibinfo {pages} {115034} (\bibinfo {year}
  {2018}),\ \Eprint{http://arxiv.org/abs/1807.10291}{arXiv:1807.10291
  [hep-ph]}%
  \bibAnnoteFile{NoStop}{Griffin:2018bjn}%
\bibitem{essig:2011nj}%
  \BibitemOpen
  \bibfield{author}{%
  \bibinfo {author} {\bibfnamefont{R.}~\bibnamefont{Essig}}, \bibinfo {author}
  {\bibfnamefont{J.}~\bibnamefont{Mardon}},\ and\ \bibinfo {author}
  {\bibfnamefont{T.}~\bibnamefont{Volansky}},\ }%
  \bibfield{journal}{%
  \Doi{10.1103/PhysRevD.85.076007}{\bibinfo {journal} {Phys. Rev.}}\ }%
  \textbf{\bibinfo {volume} {D85}},\ \bibinfo {pages} {076007} (\bibinfo {year}
  {2012}),\ \Eprint{http://arxiv.org/abs/1108.5383}{arXiv:1108.5383 [hep-ph]}%
  \bibAnnoteFile{NoStop}{essig:2011nj}%
\bibitem{Chu:2011be}%
  \BibitemOpen
  \bibfield{author}{%
  \bibinfo {author} {\bibfnamefont{X.}~\bibnamefont{Chu}}, \bibinfo {author}
  {\bibfnamefont{T.}~\bibnamefont{Hambye}},\ and\ \bibinfo {author}
  {\bibfnamefont{M.~H.~G.}\ \bibnamefont{Tytgat}},\ }%
  \bibfield{journal}{%
  \Doi{10.1088/1475-7516/2012/05/034}{\bibinfo {journal} {JCAP}}\ }%
  \textbf{\bibinfo {volume} {1205}},\ \bibinfo {pages} {034} (\bibinfo {year}
  {2012}),\ \Eprint{http://arxiv.org/abs/1112.0493}{arXiv:1112.0493 [hep-ph]}%
  \bibAnnoteFile{NoStop}{Chu:2011be}%
\bibitem{Knapen:2017xzo}%
  \BibitemOpen
  \bibfield{author}{%
  \bibinfo {author} {\bibfnamefont{S.}~\bibnamefont{Knapen}}, \bibinfo {author}
  {\bibfnamefont{T.}~\bibnamefont{Lin}},\ and\ \bibinfo {author}
  {\bibfnamefont{K.~M.}\ \bibnamefont{Zurek}},\ }%
  \bibfield{journal}{%
  \Doi{10.1103/PhysRevD.96.115021}{\bibinfo {journal} {Phys. Rev.}}\ }%
  \textbf{\bibinfo {volume} {D96}},\ \bibinfo {pages} {115021} (\bibinfo {year}
  {2017}),\ \Eprint{http://arxiv.org/abs/1709.07882}{arXiv:1709.07882
  [hep-ph]}%
  \bibAnnoteFile{NoStop}{Knapen:2017xzo}%
\bibitem{Dvorkin:2019zdi}%
  \BibitemOpen
  \bibfield{author}{%
  \bibinfo {author} {\bibfnamefont{C.}~\bibnamefont{Dvorkin}}, \bibinfo
  {author} {\bibfnamefont{T.}~\bibnamefont{Lin}},\ and\ \bibinfo {author}
  {\bibfnamefont{K.}~\bibnamefont{Schutz}}}%
   (\bibinfo {year} {2019}),\
  \Eprint{http://arxiv.org/abs/1902.08623}{arXiv:1902.08623 [hep-ph]}%
  \bibAnnoteFile{NoStop}{Dvorkin:2019zdi}%
\bibitem{Chuzhoy:2008zy}%
  \BibitemOpen
  \bibfield{author}{%
  \bibinfo {author} {\bibfnamefont{L.}~\bibnamefont{Chuzhoy}}\ and\ \bibinfo
  {author} {\bibfnamefont{E.~W.}\ \bibnamefont{Kolb}},\ }%
  \bibfield{journal}{%
  \Doi{10.1088/1475-7516/2009/07/014}{\bibinfo {journal} {JCAP}}\ }%
  \textbf{\bibinfo {volume} {0907}},\ \bibinfo {pages} {014} (\bibinfo {year}
  {2009}),\ \Eprint{http://arxiv.org/abs/0809.0436}{arXiv:0809.0436
  [astro-ph]}%
  \bibAnnoteFile{NoStop}{Chuzhoy:2008zy}%
\bibitem{McDermott:2010pa}%
  \BibitemOpen
  \bibfield{author}{%
  \bibinfo {author} {\bibfnamefont{S.~D.}\ \bibnamefont{McDermott}}, \bibinfo
  {author} {\bibfnamefont{H.-B.}\ \bibnamefont{Yu}},\ and\ \bibinfo {author}
  {\bibfnamefont{K.~M.}\ \bibnamefont{Zurek}},\ }%
  \bibfield{journal}{%
  \Doi{10.1103/PhysRevD.83.063509}{\bibinfo {journal} {Phys. Rev.}}\ }%
  \textbf{\bibinfo {volume} {D83}},\ \bibinfo {pages} {063509} (\bibinfo {year}
  {2011}),\ \Eprint{http://arxiv.org/abs/1011.2907}{arXiv:1011.2907 [hep-ph]}%
  \bibAnnoteFile{NoStop}{McDermott:2010pa}%
\bibitem{Dunsky:2018mqs}%
  \BibitemOpen
  \bibfield{author}{%
  \bibinfo {author} {\bibfnamefont{D.}~\bibnamefont{Dunsky}}, \bibinfo {author}
  {\bibfnamefont{L.~J.}\ \bibnamefont{Hall}},\ and\ \bibinfo {author}
  {\bibfnamefont{K.}~\bibnamefont{Harigaya}}}%
   (\bibinfo {year} {2018}),\
  \Eprint{http://arxiv.org/abs/1812.11116}{arXiv:1812.11116 [astro-ph.HE]}%
  \bibAnnoteFile{NoStop}{Dunsky:2018mqs}%
\bibitem{Stebbins:2019xjr}%
  \BibitemOpen
  \bibfield{author}{%
  \bibinfo {author} {\bibfnamefont{A.}~\bibnamefont{Stebbins}}\ and\ \bibinfo
  {author} {\bibfnamefont{G.}~\bibnamefont{Krnjaic}}}%
   (\bibinfo {year} {2019}),\
  \Eprint{http://arxiv.org/abs/1908.05275}{arXiv:1908.05275 [astro-ph.CO]}%
  \bibAnnoteFile{NoStop}{Stebbins:2019xjr}%
\bibitem{Freese:2012xd}%
  \BibitemOpen
  \bibfield{author}{%
  \bibinfo {author} {\bibfnamefont{K.}~\bibnamefont{Freese}}, \bibinfo {author}
  {\bibfnamefont{M.}~\bibnamefont{Lisanti}},\ and\ \bibinfo {author}
  {\bibfnamefont{C.}~\bibnamefont{Savage}},\ }%
  \bibfield{journal}{%
  \Doi{10.1103/RevModPhys.85.1561}{\bibinfo {journal} {Rev. Mod. Phys.}}\ }%
  \textbf{\bibinfo {volume} {85}},\ \bibinfo {pages} {1561} (\bibinfo {year}
  {2013}),\ \Eprint{http://arxiv.org/abs/1209.3339}{arXiv:1209.3339
  [astro-ph.CO]}%
  \bibAnnoteFile{NoStop}{Freese:2012xd}%
\bibitem{Lin:2019uvt}%
  \BibitemOpen
  \bibfield{author}{%
  \bibinfo {author} {\bibfnamefont{T.}~\bibnamefont{Lin}}}%
   (\bibinfo {year} {2019}),\
  \Eprint{http://arxiv.org/abs/1904.07915}{arXiv:1904.07915 [hep-ph]}%
  \bibAnnoteFile{NoStop}{Lin:2019uvt}%
\bibitem{Hu:2016xas}%
  \BibitemOpen
  \bibfield{author}{%
  \bibinfo {author} {\bibfnamefont{P.-K.}\ \bibnamefont{Hu}}, \bibinfo {author}
  {\bibfnamefont{A.}~\bibnamefont{Kusenko}},\ and\ \bibinfo {author}
  {\bibfnamefont{V.}~\bibnamefont{Takhistov}},\ }%
  \bibfield{journal}{%
  \Doi{10.1016/j.physletb.2017.02.035}{\bibinfo {journal} {Phys. Lett.}}\ }%
  \textbf{\bibinfo {volume} {B768}},\ \bibinfo {pages} {18} (\bibinfo {year}
  {2017}),\ \Eprint{http://arxiv.org/abs/1611.04599}{arXiv:1611.04599
  [hep-ph]}%
  \bibAnnoteFile{NoStop}{Hu:2016xas}%
\bibitem{Cerdonio:1997hz}%
  \BibitemOpen
  \bibfield{author}{%
  \bibinfo {author} {\bibfnamefont{M.}~\bibnamefont{Cerdonio}} \emph{et~al.},\
  }%
  \bibfield{booktitle}{%
  \emph{\bibinfo {booktitle} {{1st International LISA Symposium on
  Gravitational Waves Oxfordshire, England, July 9-12, 1996}}},\ }%
  \bibfield{journal}{%
  \Doi{10.1088/0264-9381/14/6/016}{\bibinfo {journal} {Class. Quant. Grav.}}\
  }%
  \textbf{\bibinfo {volume} {14}},\ \bibinfo {pages} {1491} (\bibinfo {year}
  {1997})%
  \bibAnnoteFile{NoStop}{Cerdonio:1997hz}%
\bibitem{McAllister:2018ndu}%
  \BibitemOpen
  \bibfield{author}{%
  \bibinfo {author} {\bibfnamefont{B.~T.}\ \bibnamefont{McAllister}}, \bibinfo
  {author} {\bibfnamefont{M.}~\bibnamefont{Goryachev}}, \bibinfo {author}
  {\bibfnamefont{J.}~\bibnamefont{Bourhill}}, \bibinfo {author}
  {\bibfnamefont{E.~N.}\ \bibnamefont{Ivanov}},\ and\ \bibinfo {author}
  {\bibfnamefont{M.~E.}\ \bibnamefont{Tobar}}}%
   (\bibinfo {year} {2018}),\
  \Eprint{http://arxiv.org/abs/1803.07755}{arXiv:1803.07755 [physics.ins-det]}%
  \bibAnnoteFile{NoStop}{McAllister:2018ndu}%
\bibitem{Ouellet:2018nfr}%
  \BibitemOpen
  \bibfield{author}{%
  \bibinfo {author} {\bibfnamefont{J.}~\bibnamefont{Ouellet}}\ and\ \bibinfo
  {author} {\bibfnamefont{Z.}~\bibnamefont{Bogorad}},\ }%
  \bibfield{journal}{%
  \Doi{10.1103/PhysRevD.99.055010}{\bibinfo {journal} {Phys. Rev.}}\ }%
  \textbf{\bibinfo {volume} {D99}},\ \bibinfo {pages} {055010} (\bibinfo {year}
  {2019}),\ \Eprint{http://arxiv.org/abs/1809.10709}{arXiv:1809.10709
  [hep-ph]}%
  \bibAnnoteFile{NoStop}{Ouellet:2018nfr}%
\bibitem{Beutter:2018xfx}%
  \BibitemOpen
  \bibfield{author}{%
  \bibinfo {author} {\bibfnamefont{M.}~\bibnamefont{Beutter}}, \bibinfo
  {author} {\bibfnamefont{A.}~\bibnamefont{Pargner}}, \bibinfo {author}
  {\bibfnamefont{T.}~\bibnamefont{Schwetz}},\ and\ \bibinfo {author}
  {\bibfnamefont{E.}~\bibnamefont{Todarello}},\ }%
  \bibfield{journal}{%
  \Doi{10.1088/1475-7516/2019/02/026}{\bibinfo {journal} {JCAP}}\ }%
  \textbf{\bibinfo {volume} {1902}},\ \bibinfo {pages} {026} (\bibinfo {year}
  {2019}),\ \Eprint{http://arxiv.org/abs/1812.05487}{arXiv:1812.05487
  [hep-ph]}%
  \bibAnnoteFile{NoStop}{Beutter:2018xfx}%
\bibitem{Nyquist:1928zz}%
  \BibitemOpen
  \bibfield{author}{%
  \bibinfo {author} {\bibfnamefont{H.}~\bibnamefont{Nyquist}},\ }%
  \bibfield{journal}{%
  \Doi{10.1103/PhysRev.32.110}{\bibinfo {journal} {Phys. Rev.}}\ }%
  \textbf{\bibinfo {volume} {32}},\ \bibinfo {pages} {110} (\bibinfo {year}
  {1928})%
  \bibAnnoteFile{NoStop}{Nyquist:1928zz}%
\bibitem{Budker:2013hfa}%
  \BibitemOpen
  \bibfield{author}{%
  \bibinfo {author} {\bibfnamefont{D.}~\bibnamefont{Budker}}, \bibinfo {author}
  {\bibfnamefont{P.~W.}\ \bibnamefont{Graham}}, \bibinfo {author}
  {\bibfnamefont{M.}~\bibnamefont{Ledbetter}}, \bibinfo {author}
  {\bibfnamefont{S.}~\bibnamefont{Rajendran}},\ and\ \bibinfo {author}
  {\bibfnamefont{A.}~\bibnamefont{Sushkov}},\ }%
  \bibfield{journal}{%
  \Doi{10.1103/PhysRevX.4.021030}{\bibinfo {journal} {Phys. Rev.}}\ }%
  \textbf{\bibinfo {volume} {X4}},\ \bibinfo {pages} {021030} (\bibinfo {year}
  {2014}),\ \Eprint{http://arxiv.org/abs/1306.6089}{arXiv:1306.6089 [hep-ph]}%
  \bibAnnoteFile{NoStop}{Budker:2013hfa}%
\bibitem{Essig:2012yx}%
  \BibitemOpen
  \bibfield{author}{%
  \bibinfo {author} {\bibfnamefont{R.}~\bibnamefont{Essig}}, \bibinfo {author}
  {\bibfnamefont{A.}~\bibnamefont{Manalaysay}}, \bibinfo {author}
  {\bibfnamefont{J.}~\bibnamefont{Mardon}}, \bibinfo {author}
  {\bibfnamefont{P.}~\bibnamefont{Sorensen}},\ and\ \bibinfo {author}
  {\bibfnamefont{T.}~\bibnamefont{Volansky}},\ }%
  \bibfield{journal}{%
  \Doi{10.1103/PhysRevLett.109.021301}{\bibinfo {journal} {Phys. Rev. Lett.}}\
  }%
  \textbf{\bibinfo {volume} {109}},\ \bibinfo {pages} {021301} (\bibinfo {year}
  {2012}),\ \Eprint{http://arxiv.org/abs/1206.2644}{arXiv:1206.2644
  [astro-ph.CO]}%
  \bibAnnoteFile{NoStop}{Essig:2012yx}%
\bibitem{Essig:2017kqs}%
  \BibitemOpen
  \bibfield{author}{%
  \bibinfo {author} {\bibfnamefont{R.}~\bibnamefont{Essig}}, \bibinfo {author}
  {\bibfnamefont{T.}~\bibnamefont{Volansky}},\ and\ \bibinfo {author}
  {\bibfnamefont{T.-T.}\ \bibnamefont{Yu}},\ }%
  \bibfield{journal}{%
  \Doi{10.1103/PhysRevD.96.043017}{\bibinfo {journal} {Phys. Rev.}}\ }%
  \textbf{\bibinfo {volume} {D96}},\ \bibinfo {pages} {043017} (\bibinfo {year}
  {2017}),\ \Eprint{http://arxiv.org/abs/1703.00910}{arXiv:1703.00910
  [hep-ph]}%
  \bibAnnoteFile{NoStop}{Essig:2017kqs}%
\bibitem{Vogel:2013raa}%
  \BibitemOpen
  \bibfield{author}{%
  \bibinfo {author} {\bibfnamefont{H.}~\bibnamefont{Vogel}}\ and\ \bibinfo
  {author} {\bibfnamefont{J.}~\bibnamefont{Redondo}},\ }%
  \bibfield{journal}{%
  \Doi{10.1088/1475-7516/2014/02/029}{\bibinfo {journal} {JCAP}}\ }%
  \textbf{\bibinfo {volume} {1402}},\ \bibinfo {pages} {029} (\bibinfo {year}
  {2014}),\ \Eprint{http://arxiv.org/abs/1311.2600}{arXiv:1311.2600 [hep-ph]}%
  \bibAnnoteFile{NoStop}{Vogel:2013raa}%
\bibitem{Chang:2018rso}%
  \BibitemOpen
  \bibfield{author}{%
  \bibinfo {author} {\bibfnamefont{J.~H.}\ \bibnamefont{Chang}}, \bibinfo
  {author} {\bibfnamefont{R.}~\bibnamefont{Essig}},\ and\ \bibinfo {author}
  {\bibfnamefont{S.~D.}\ \bibnamefont{McDermott}},\ }%
  \bibfield{journal}{%
  \Doi{10.1007/JHEP09(2018)051}{\bibinfo {journal} {JHEP}}\ }%
  \textbf{\bibinfo {volume} {09}},\ \bibinfo {pages} {051} (\bibinfo {year}
  {2018}),\ \Eprint{http://arxiv.org/abs/1803.00993}{arXiv:1803.00993
  [hep-ph]}%
  \bibAnnoteFile{NoStop}{Chang:2018rso}%
\bibitem{Kovetz:2018zan}%
  \BibitemOpen
  \bibfield{author}{%
  \bibinfo {author} {\bibfnamefont{E.~D.}\ \bibnamefont{Kovetz}}, \bibinfo
  {author} {\bibfnamefont{V.}~\bibnamefont{Poulin}}, \bibinfo {author}
  {\bibfnamefont{V.}~\bibnamefont{Gluscevic}}, \bibinfo {author}
  {\bibfnamefont{K.~K.}\ \bibnamefont{Boddy}}, \bibinfo {author}
  {\bibfnamefont{R.}~\bibnamefont{Barkana}},\ and\ \bibinfo {author}
  {\bibfnamefont{M.}~\bibnamefont{Kamionkowski}},\ }%
  \bibfield{journal}{%
  \Doi{10.1103/PhysRevD.98.103529}{\bibinfo {journal} {Phys. Rev.}}\ }%
  \textbf{\bibinfo {volume} {D98}},\ \bibinfo {pages} {103529} (\bibinfo {year}
  {2018}),\ \Eprint{http://arxiv.org/abs/1807.11482}{arXiv:1807.11482
  [astro-ph.CO]}%
  \bibAnnoteFile{NoStop}{Kovetz:2018zan}%
\bibitem{Williams:1971ms}%
  \BibitemOpen
  \bibfield{author}{%
  \bibinfo {author} {\bibfnamefont{E.~R.}\ \bibnamefont{Williams}}, \bibinfo
  {author} {\bibfnamefont{J.~E.}\ \bibnamefont{Faller}},\ and\ \bibinfo
  {author} {\bibfnamefont{H.~A.}\ \bibnamefont{Hill}},\ }%
  \bibfield{journal}{%
  \Doi{10.1103/PhysRevLett.26.721}{\bibinfo {journal} {Phys. Rev. Lett.}}\ }%
  \textbf{\bibinfo {volume} {26}},\ \bibinfo {pages} {721} (\bibinfo {year}
  {1971})%
  \bibAnnoteFile{NoStop}{Williams:1971ms}%
\bibitem{Baggio:2005xp}%
  \BibitemOpen
  \bibfield{author}{%
  \bibinfo {author} {\bibfnamefont{L.}~\bibnamefont{Baggio}} \emph{et~al.},\ }%
  \bibfield{journal}{%
  \Doi{10.1103/PhysRevLett.94.241101}{\bibinfo {journal} {Phys. Rev. Lett.}}\
  }%
  \textbf{\bibinfo {volume} {94}},\ \bibinfo {pages} {241101} (\bibinfo {year}
  {2005}),\ \Eprint{http://arxiv.org/abs/gr-qc/0502101}{arXiv:gr-qc/0502101
  [gr-qc]}%
  \bibAnnoteFile{NoStop}{Baggio:2005xp}%
\bibitem{Bonaldi:1998gcg}%
  \BibitemOpen
  \bibfield{author}{%
  \bibinfo {author} {\bibfnamefont{M.}~\bibnamefont{Bonaldi}}, \bibinfo
  {author} {\bibfnamefont{P.}~\bibnamefont{Falferi}}, \bibinfo {author}
  {\bibfnamefont{R.}~\bibnamefont{Dolesi}}, \bibinfo {author}
  {\bibfnamefont{M.}~\bibnamefont{Cerdonio}},\ and\ \bibinfo {author}
  {\bibfnamefont{S.}~\bibnamefont{Vitale}},\ }%
  \bibfield{journal}{%
  \Doi{10.1063/1.1149166}{\bibinfo {journal} {Rev. Sci. Instrum.}}\ }%
  \textbf{\bibinfo {volume} {69}},\ \bibinfo {pages} {3690} (\bibinfo {year}
  {1998})%
  \bibAnnoteFile{NoStop}{Bonaldi:1998gcg}%
\bibitem{Bonaldi:1999mvu}%
  \BibitemOpen
  \bibfield{author}{%
  \bibinfo {author} {\bibfnamefont{M.}~\bibnamefont{Bonaldi}}, \bibinfo
  {author} {\bibfnamefont{P.}~\bibnamefont{Falferi}}, \bibinfo {author}
  {\bibfnamefont{M.}~\bibnamefont{Cerdonio}}, \bibinfo {author}
  {\bibfnamefont{A.}~\bibnamefont{Vinante}}, \bibinfo {author}
  {\bibfnamefont{R.}~\bibnamefont{Dolesi}},\ and\ \bibinfo {author}
  {\bibfnamefont{S.}~\bibnamefont{Vitale}},\ }%
  \bibfield{journal}{%
  \Doi{10.1063/1.1149679}{\bibinfo {journal} {Rev. Sci. Instrum.}}\ }%
  \textbf{\bibinfo {volume} {70}},\ \bibinfo {pages} {1851} (\bibinfo {year}
  {1999})%
  \bibAnnoteFile{NoStop}{Bonaldi:1999mvu}%
\bibitem{Heavner_2014}%
  \BibitemOpen
  \bibfield{author}{%
  \bibinfo {author} {\bibfnamefont{T.~P.}\ \bibnamefont{Heavner}}, \bibinfo
  {author} {\bibfnamefont{E.~A.}\ \bibnamefont{Donley}}, \bibinfo {author}
  {\bibfnamefont{F.}~\bibnamefont{Levi}}, \bibinfo {author}
  {\bibfnamefont{G.}~\bibnamefont{Costanzo}}, \bibinfo {author}
  {\bibfnamefont{T.~E.}\ \bibnamefont{Parker}}, \bibinfo {author}
  {\bibfnamefont{J.~H.}\ \bibnamefont{Shirley}}, \bibinfo {author}
  {\bibfnamefont{N.}~\bibnamefont{Ashby}}, \bibinfo {author}
  {\bibfnamefont{S.}~\bibnamefont{Barlow}},\ and\ \bibinfo {author}
  {\bibfnamefont{S.~R.}\ \bibnamefont{Jefferts}},\ }%
  \bibfield{journal}{%
  \Doi{10.1088/0026-1394/51/3/174}{\bibinfo {journal} {Metrologia}}\ }%
  \textbf{\bibinfo {volume} {51}},\ \bibinfo {pages} {174} (\bibinfo {month}
  {may}\ \bibinfo {year} {2014}),\
  \url{https://doi.org/10.1088%2F0026-1394%2F51%2F3%2F174}%
  \bibAnnoteFile{NoStop}{Heavner_2014}%
\bibitem{Necib:2018iwb}%
  \BibitemOpen
  \bibfield{author}{%
  \bibinfo {author} {\bibfnamefont{L.}~\bibnamefont{Necib}}, \bibinfo {author}
  {\bibfnamefont{M.}~\bibnamefont{Lisanti}},\ and\ \bibinfo {author}
  {\bibfnamefont{V.}~\bibnamefont{Belokurov}}}%
   (\bibinfo {year} {2018}),\ \doi{\bibinfo {doi} {10.3847/1538-4357/ab095b}},\
  \Eprint{http://arxiv.org/abs/1807.02519}{arXiv:1807.02519 [astro-ph.GA]}%
  \bibAnnoteFile{NoStop}{Necib:2018iwb}%
\bibitem{Mitra:2006ds}%
  \BibitemOpen
  \bibfield{author}{%
  \bibinfo {author} {\bibfnamefont{S.}~\bibnamefont{Mitra}},\ }%
  \bibfield{journal}{%
  \Doi{10.1103/PhysRevD.74.043532}{\bibinfo {journal} {Phys. Rev.}}\ }%
  \textbf{\bibinfo {volume} {D74}},\ \bibinfo {pages} {043532} (\bibinfo {year}
  {2006}),\
  \Eprint{http://arxiv.org/abs/astro-ph/0605369}{arXiv:astro-ph/0605369
  [astro-ph]}%
  \bibAnnoteFile{NoStop}{Mitra:2006ds}%
\bibitem{Geraci:2017bmq}%
  \BibitemOpen
  \bibfield{author}{%
  \bibinfo {author} {\bibfnamefont{A.~A.}\ \bibnamefont{Geraci}} \emph{et~al.}
  (\bibinfo {collaboration} {ARIADNE}),\ }%
  \bibfield{booktitle}{%
  \emph{\bibinfo {booktitle} {{Proceedings, 2nd Workshop on Microwave Cavities
  and Detectors for Axion Research: Livermore, California, USA, January 10-13,
  2017}}},\ }%
  \bibfield{journal}{%
  \Doi{10.1007/978-3-319-92726-8_18}{\bibinfo {journal} {Springer Proc.
  Phys.}}\ }%
  \textbf{\bibinfo {volume} {211}},\ \bibinfo {pages} {151} (\bibinfo {year}
  {2018}),\ \Eprint{http://arxiv.org/abs/1710.05413}{arXiv:1710.05413
  [astro-ph.IM]}%
  \bibAnnoteFile{NoStop}{Geraci:2017bmq}%
\end{thebibliography}%


\clearpage
\newpage
\maketitle
\onecolumngrid
\begin{center}
\textbf{\large Direct Deflection of Particle Dark Matter} \\ 
\vspace{0.05in}
{ \it \large Supplementary Material}\\ 
\vspace{0.05in}
{}
{Asher Berlin, Raffaele Tito D'Agnolo, Sebastian A. R. Ellis, Philip Schuster, Natalia Toro}

\end{center}
\setcounter{equation}{0}
\setcounter{figure}{0}
\setcounter{table}{0}
\setcounter{section}{1}
\renewcommand{\theequation}{S\arabic{equation}}
\renewcommand{\thefigure}{S\arabic{figure}}
\renewcommand{\thetable}{S\arabic{table}}
\newcommand\ptwiddle[1]{\mathord{\mathop{#1}\limits^{\scriptscriptstyle(\sim)}}}

In this Supplementary Material, we first give a detailed derivation of the DM charge and current densities. In the later sections, we investigate the DM-induced electromagnetic fields for massive dark photons, and show that this reduces to the simple massless case when the dark photon is much lighter than the inverse geometric size of the experimental setup in Fig.~\ref{fig:setup}.

\section*{Induced Dark Matter Charge and Current Densities}

\subsubsection{General Analysis}

The DM plasma is assumed to consist of particles that are either positively or negatively charged under electromagnetism, with both species contributing equally to the total cosmological DM number density ($n_\x$).  We further assume that, in the absence of external fields, the two species have identical and spatially uniform velocity phase-space densities 
\be
\label{ffreeA}
f_j^{\text{free}}(\xv, \vv) \equiv \frac{1}{2} \, n_\x \, f(\vv)
~,
\ee
where $j=0  (1)$ correspond to positive (negative) charges, respectively, and $f(\vv)$ is the unit-normalized DM velocity distribution.  

DM charge and current densities arise because external fields deflect positive and negative charges differently.  Since the DM plasma is essentially non-interacting, we can account for these effects by writing the initial density of each species (at an early time, $t_0$, when they are unaffected by external fields) as a sum over particle positions and velocities:
\be
\label{eq:ffreeB}
f_j(\xv, \vv, t_0) \equiv  \frac{1}{2} \, n_\x  \, \int d^3\xv_i  \, d^3\vv_i  ~ f(\vv_i) ~ \delta^{(3)}(\xv-\xv_i) ~ \delta^{(3)}(\vv-\vv_i)
~.
\ee
The above expression is, of course, related to Eq.~(\ref{ffreeA}) by a trivial insertion of unity, but can be physically interpreted as populating the DM plasma with test particles at positions $\xv_i$ with velocity $\vv_i$.  In the integrand of Eq.~(\ref{eq:ffreeB}),  we promote $\xv_i$ and $\vv_i$ to time-dependent phase-space coordinates, $\xv_\defl(t; \xv_i, \vv_i)$ and $\vv_\defl(t; \xv_i, \vv_i)$, that follow a classical equation of motion with boundary condition 
$\xv_\defl(t_0; \xv_i, \vv_i) = \xv_i$ and $\vv_\defl(t_0; \xv_i, \vv_i) = \vv_i$. We can then derive expressions for the time-evolved phase-space density, $f_j$, in the presence of arbitrary background electromagnetic fields. Although Eq.~(\ref{eq:ffreeB}) relates the phase space density to an integral over classical particle trajectories, it reflects the continuum limit, i.e., the expected density averaged over Poission fluctuations (which, as discussed in the text, are acceptably small when the phase-space density of Eq.~(\ref{eq:ffreeB}) is integrated over physically relevant regions).

Summing over positively and negatively charged DM species, the DM charge density ($\rho_\x$) at time $t$ can be expressed as
\begin{align}
\label{eq:rho1}
\rho_\x (\xv , t)  &= e \qeff \, \sum\limits_{j = 0}^1 (-1)^j \, \int d^3\vv ~ f_j(\xv , \vv, t)  
\nl
&=  \frac{1}{2} \, e \qeff ~ n_\x  \, \sum\limits_{j = 0}^1 (-1)^j \, \int d^3\xv_i \, d^3\vv_i ~ f(\vv_i) ~ \delta^{(3)}(\xv-\xv_\defl(t;\xv_i,\vv_i))
~,
\end{align}
and similarly for the current density, 
\begin{align}
\label{eq:j1}
\jx (\xv , t)  &=  e \qeff \, \sum\limits_{j = 0}^1 (-1)^j \, \int d^3\vv ~ f_j(\xv , \vv, t) ~ \vv
\nl
&= \frac{1}{2} \, e \qeff ~ n_\x \, \sum\limits_{j = 0}^1 (-1)^j \, \int d^3\xv_i \, d^3\vv_i ~ f(\vv_i) ~ \vv_\defl(t;\xv_i,\vv_i) ~ \delta^{(3)}(\xv-\xv_\defl(t;\xv_i,\vv_i))
~.
\end{align}

We will be interested in the case that the motion of millicharged DM is weakly affected by the deflector electromagnetic fields. In this case, we can decompose the deflected DM trajectory as
\be 
\label{freeExpansion}
\xv_\defl \equiv \xv_\text{free} + \Delta \xv_\defl
~,~  
\vv_\defl \equiv \vv_\text{free} + \Delta \vv_\defl
~,
\ee
where we have suppressed dependence on $t$, $\xv_i$, and $\vv_i$, and the free-particle trajectory is given by 
\be
\xv_\text{free}(t) \equiv \xv_i + \vv (t-t_0)
~,~
 \vv_\text{free}(t) \equiv \vv_i
 ~.
\ee
The small position and velocity deflections, $\Delta \xv_\defl $ and $\Delta \vv_\defl$, can be approximated by integrating the electromagnetic deflector force along the \emph{free} equation of motion. Neglecting magnetic field effects, the deflection in position can be written in terms of the deflector electric field, $\E_\defl (\xv, t) = \E_\defl (\xv) \, e^{i \omega t}$, as
\begin{align}
\label{eq:Lorentz}
\Delta \xv_\defl (t) &\simeq (-1)^j ~ \frac{e \qeff}{m_\x} ~~ \iint\limits_{t_0 < \tp < \tpp < t} \, d \tp \, d \tpp ~ \E_\defl (\xv_\text{free}(\tp)) \, e^{i \omega \tp} \\
~
 &\simeq (-1)^j ~ \frac{e \qeff}{m_\x} ~~ \int_0^\infty d\tau ~ \tau ~ \E_\defl (\xv_\text{free}(t) - \vv_i \tau ) \, e^{i \omega (t-\tau)},
\end{align}
where in the second line we evaluated the $\tpp$ integral, took $t_0 \to - \infty$, and changed variables in the remaining integral from $\tp$ to $\tau \equiv t-\tp$. Below, we drop the subscript on $\vv_i$, such that $\vv$ refers to the time-independent free-particle velocity $\vv_\text{free} = \vv_i$.

Inserting Eq.~(\ref{freeExpansion}) into Eq.~(\ref{eq:rho1}) and expanding to leading order in small deflections ($|\Delta \xv_\defl| \ll |\xv_\text{free}|$), we find 
\begin{align}
\label{eq:rho3}
\rho_\x(\xv,t) &\simeq \frac{1}{2} \, e \qeff ~ n_\x \, \sum\limits_{j = 0}^1 (-1)^j \int d^3 \xv_i \, d^3 \vv ~ f(\vv) \, \left( 1 + \Delta \xv_\defl \cdot \lap_{\xv_i} \right) \delta^{(3)} (\xv - \xv_\text{free} )
\nl
&\simeq - \frac{1}{2} \, e \qeff ~ n_\x \, \sum\limits_{j = 0}^1 (-1)^j \int d^3 \vv ~ f(\vv) \, \left( \lap_{\xv_i} \cdot \Delta \xv_\defl \right) \Bigg|_{\xv_i = \xv - \vv (t-t_0)}
~,
\end{align}
where in the second line we have dropped terms that cancel in the sum over positively and negatively charged species and integrated by parts. From Eq.~(\ref{eq:Lorentz}) and using Gauss's law, we arrive at the following compact expression for the induced DM charge density:
\be
\label{eq:rho4}
\rho_\x (\xv , t) \simeq - \frac{(e \qeff)^2 \rhodm}{m_\x^2} ~ e^{i \omega t} \, \int d^3 \vv ~ f(\vv)  \int_0^\infty d \tau ~ \tau \, \rho_\defl (\xv - \vv \, \tau) \, e^{-i \omega \tau}
~,
\ee
where $\rho_\defl$ is the charge density of the deflector and $\rhodm \simeq m_\x \, n_\x$ is the ambient DM \emph{energy} density.

We now make a final change of variables to $\xv^\p \equiv \xv - \vv \tau$ and $v \equiv |\vv|$, whose Jacobian is
\be
d^3 \vv ~ d \tau = (v \, \tau^2)^{-1} \, d^3 \xv^\p ~ d v
~.
\ee
Switching to this new basis, Eq.~(\ref{eq:rho4}) becomes
\be
\label{eq:rho5}
\eqbox{
\rho_\x (\xv , t) \simeq - \frac{(e \qeff)^2 \rhodm}{m_\x^2} ~ e^{i \omega t} \, \int d v ~ d^3 \xv^\p ~ f(v \, \hat{\vv})  \, \frac{\rho_\defl (\xv^\p)}{|\xv - \xv^\p|} \, e^{-i \omega |\xv - \xv^\p| / v}
}
~,
\ee
where the unit-vector, $\hat{\vv}$, is defined to be a function of $\xv$ and $\xv^\p$,
\be
\hat{\vv} \equiv \frac{\xv - \xv^\p}{|\xv - \xv^\p|}
~.
\ee

The procedure for calculating induced DM current densities ($\jx$) follows in a similar manner from Eq.~(\ref{eq:j1}).   In addition to the deformation of the trajectory in the position $\delta$-function, $\jx$ receives a second contribution at leading order in the small deflecting force, due to the explicit $\vv$-dependence of the integrand. In particular, the DM velocity receives a leading order deflection correction 
\begin{align}
\Delta \vv_\defl (t) 
&\simeq (-1)^j ~ \frac{e \qeff}{m_\x} \, \int_{t_0}^t d \tp ~ \E_\defl (\xv_\text{free}(\tp)) \, e^{i \omega \tp}
\nl
&\simeq (-1)^j ~ \frac{e \qeff}{m_\x} \, \int_0^\infty d \tau ~ \E_\defl (\xv_\text{free}(t)-\vv \tau) \, e^{i \omega (t-\tau)}
~.
\end{align}
Combining both the $\Delta \xv_\defl$ and $\Delta \vv_\defl$ contributions, we find 
\be
\label{eq:j2}
\eqbox{
\jx (\xv , t) \simeq \frac{(e \qeff)^2 \rhodm}{m_\x^2} ~ e^{i \omega t} \, \int d v ~ d^3 \xv^\p ~ f(v \, \hat{\vv})  \, \frac{v}{|\xv - \xv^\p|^2} \, \Big( \E_\defl(\xv^\p) - (\xv - \xv^\p) \rho_\defl (\xv^\p) \Big) \, e^{-i \omega |\xv - \xv^\p| / v}
}
~.
\ee
Taking the divergence of Eq.~(\ref{eq:j2}) and using the identity
\be
\lap \cdot \left( \frac{f(v \, \hat{\vv})}{|\xv-\xv^\p|^2} \, (\xv - \xv^\p) \right) = \frac{f(v \, \hat{\vv})}{|\xv - \xv^\p|^2}
~,
\ee
it is straightforward to show that Eqs.~(\ref{eq:rho5}) and (\ref{eq:j2}) are related by continuity ($\partial_t \rho_\x + \lap \cdot \jx = 0$). In the calculations that follow, we will work in the quasi-static limit, in which case $e^{- i \omega |\xv - \xv^\p| / v} \to 1$ in the integrands of Eqs.~(\ref{eq:rho5}) and (\ref{eq:j2}).

\subsubsection{Debye Screening}

As a concrete example, let us focus on the static ($\omega = 0$) limit with an isotropic Maxwellian distribution, $f(\vv) \propto e^{-|\vv|^2/v_0^2}.$
 In this case, $f(\vv)$ is independent of the direction of $\vv$ and hence $f(v \, \hat{\vv}) = f(v)$. As a result, the integrals over $v$ and $\xv^\p$ in Eq.~(\ref{eq:rho5}) factorize. Recognizing the $\xv^\p$ integral  in Eq.~(\ref{eq:rho5}) as the definition of the deflector electric potential ($\phi_\defl$) and explicitly evaluating the integral over $v$, we find
\be
\label{eq:debye}
\rho_\x (\xv) \simeq - \frac{(e \qeff)^2 \rhodm}{m_\x} ~ \frac{\phi_\defl (\xv)}{T}
~,
\ee
where we have identified the effective DM temperature as
\be
T \equiv (m_\x/3) \, \langle v^2 \rangle \simeq (m_\x / 2) v_0^2
~.
\ee
Eq.~(\ref{eq:debye}) corresponds to the standard result of Debye screening of an electric potential by a weakly-coupled plasma.  We note that \emph{any} isotropic velocity distribution would lead to a similar form of $\rho_\x$, proportional to the electrostatic potential $\phi_\defl$ (albeit with a different temperature).  As we will see below, anisotropy of $f(\vv)$ (for example, arising from a DM wind) and the resulting non-trivial $\hat \vv$ dependence is essential to achieving a non-zero $\rho_\x$ in regions of vanishing electric potential.  

\subsubsection{Far-Field Limit}

In general, Eqs.~(\ref{eq:rho5}) and (\ref{eq:j2}) 
must be solved numerically. However, far outside of the deflector region ($|\xv| \gg R$), Eq.~(\ref{eq:rho5}) can be expressed as a sum over multipole and trace moments of the deflector charge distribution.  To do this, we first rewrite Eq.~(\ref{eq:rho5}) as
\be
\label{eq:rho6}
\rho_\x (\xv , t) \simeq - \frac{(e \qeff)^2 \rhodm}{m_\x^2} ~ e^{i \omega t} \, \int d^3 \xv^\p ~ \rho_\defl (\xv^\p)  ~ G (\xv - \xv^\p)
~,
\ee
where we have defined
\be
\label{eq:Gdef}
G (\yv) \equiv \int dv\, \frac{f(v \, \hat{\yv})}{|\yv|}
~.
\ee
Note that $G(\xv - \xv^\p)$ is smooth about $\xv^\p = 0$ and can be replaced in Eq.~(\ref{eq:rho6}) by its Taylor expansion in $\xv^\p$.  For each term in the Taylor series, the $\xv^\p$ integral reduces to a moment of the deflector charge distribution.  Thus, we have 
\be
\label{eq:multipole0}
\rho_\x (\xv, t) \simeq - \frac{(e \qeff)^2 \rhodm}{m_\x^2} ~ e^{i \omega t} \, \left( \rho_\x^{(1)} + \rho_\x^{(2)} + \rho_\x^{(3)} + \cdots \right)
~
\ee
where 
\begin{align}
\label{eq:multipole}
\rho_\x^{(1)} &\equiv G(\xv-\xv^\p)\big|_{\xv^\p = 0}  ~ \int d^3 \xv^\p ~ \rho_\defl(\xv^\p) = Q_\defl \, G(\xv)
\nl
\rho_\x^{(2)} &\equiv  \lap_i^\p G(\xv-\xv^\p) \big|_{\xv^\p = 0} ~ \int d^3 \xv^\p ~ x^{\p i } \,  \rho_\defl(\xv^\p) = - p_\defl^i \,  \lap_i G(\xv)
\nl
\rho_\x^{(3)} &\equiv \frac{1}{2} \, \lap_i^\p \lap_j^\p G(\xv-\xv^\p) \big|_{\xv^\p = 0} ~ \int d^3 \xv^\p ~ x^{\p i } x^{\p j } \,  \rho_\defl(\xv^\p) = \frac{1}{6} \left( Q_\defl^{i j}  \, \lap_i \lap_j G(\xv) + \mathcal{R}_\defl^2 \, \lap^{2} G(\xv) \right)
~.
\end{align}
Above, a sum over $i,j = 1,2,3$ is implied and $\lap^\p$ corresponds to differentiation with respect to $\xv^\p$.  
In the second equality of each line of Eq.~(\ref{eq:multipole}), we have rewritten the integral over the deflector volume in terms of the total charge ($Q_\defl$), the dipole moment ($p_\defl^i$), the quadrupole moment ($Q_\defl^{ij}$), and the charge-radius-squared ($\mathcal{R}_\defl^2$) of the deflection region, and replaced $\lap^{\p}  \rightarrow -\lap$ since $G (\xv - \xv^\p)$ is a function of $\xv - \xv^\p$. The term $\rho_\x^{(j)}$ scales as $1/|\xv|^j$.

\begin{figure*}[t!]
\includegraphics[width = 0.49\textwidth]{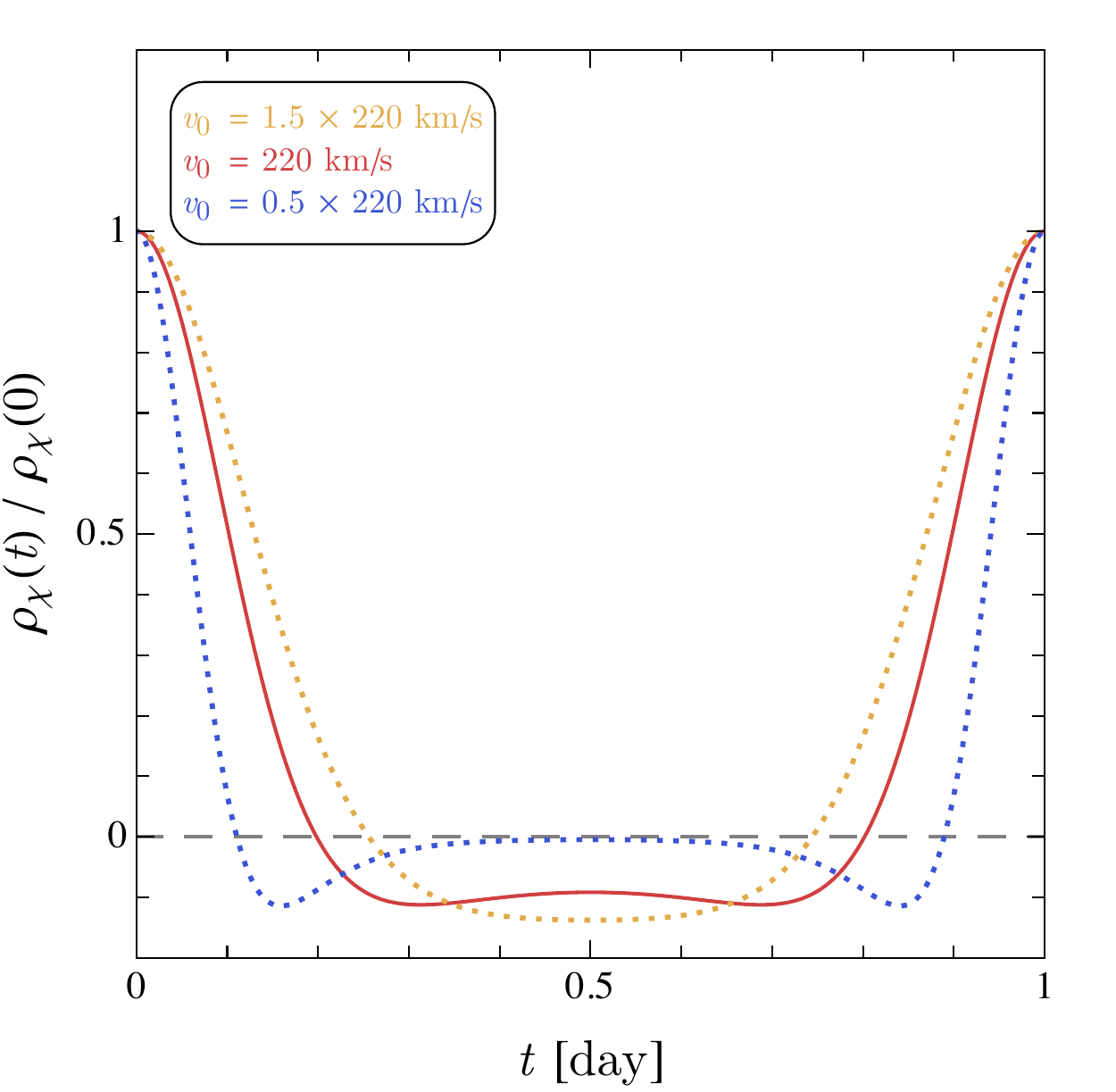}
\includegraphics[width = 0.49\textwidth]{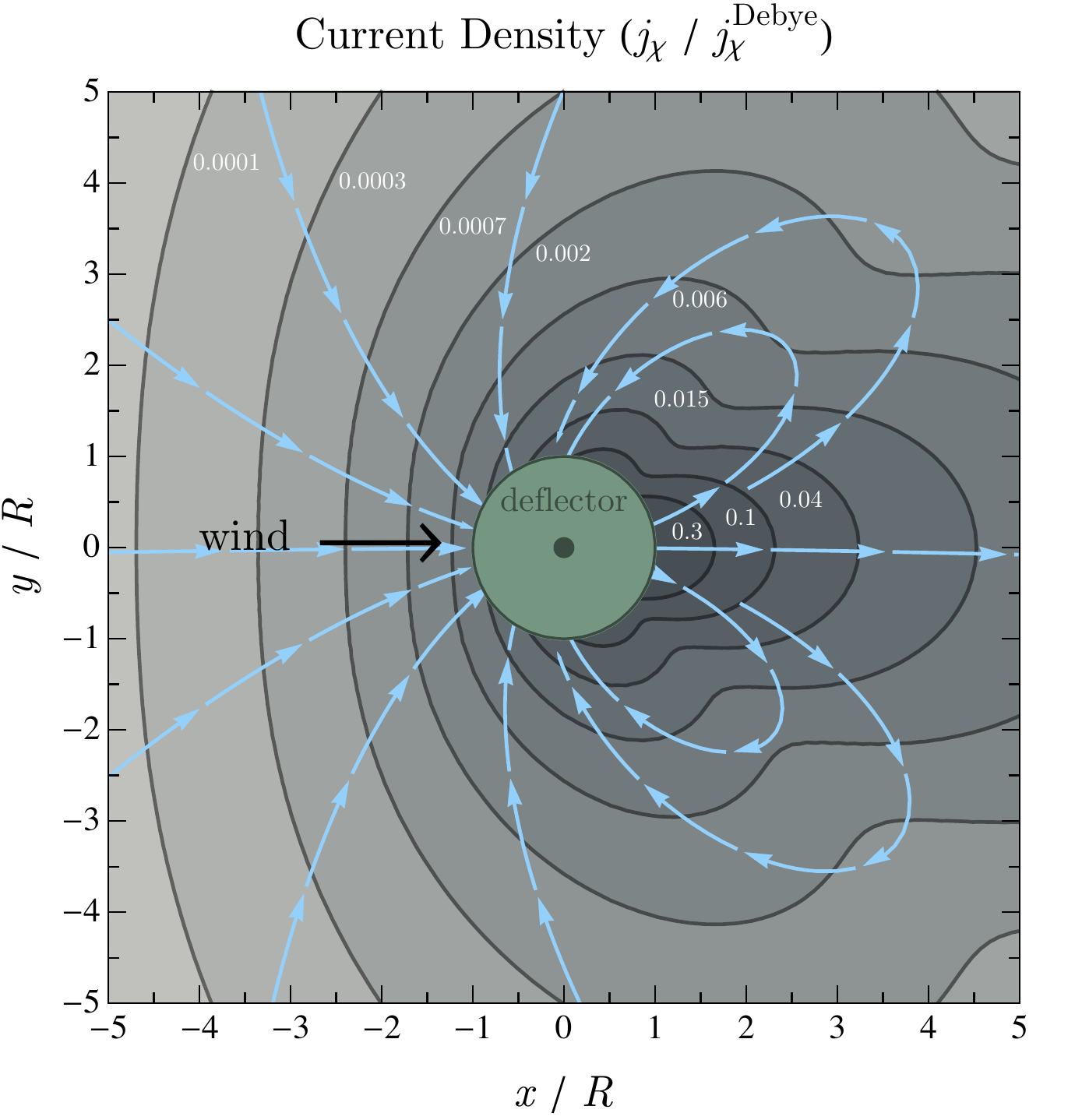}
\vspace{-0.2cm}
\caption{(Left) The daily modulation of the dark matter charge density amplitude, $\rho_\x$, induced by a deflector consisting of a point charge inside a spherical shield and normalized by its maximum value at time $t = 0$. We assume a shifted-Maxwellian velocity distribution (see Eq.~(\ref{eq:fv})) and vary the velocity dispersion, $v_0$. The solid red curve corresponds to the commonly accepted value of the dark matter velocity dispersion. For the dotted lines, the dispersion velocity  is rescaled to values 50\% larger and smaller in order to show the signal's sensitivity to variations in $v_0$. (Right) The time-independent amplitude of the dark matter current density, $\jv_\x$, in the $x-y$ plane, induced by a deflector consisting of a point charge inside a spherical shield of radius $R$ (as given by the far-field approximation of Eq.~(\ref{eq:multipole2})). The direction of the dark matter wind is along the positive $x$ direction, as specified by the black arrow. 
The dark gray contours correspond to values of $|\jv_\x|$ when normalized by the dimensionful quantity $|j_\x^\text{Debye}|$ (see Eq.~(\ref{eq:jdebye})). The light blue arrows refer to the direction of $\jv_\x / j_\x^\text{Debye}$ projected on the $x-y$ plane.
}
\label{fig:angleandcurrent}
\vspace{-0.5cm}
\end{figure*}

The moment expansion introduced above simplifies dramatically when we specialize to configurations where the deflection region is bounded by a grounded conductor, as in Fig.~\ref{fig:setup}.  An ideal grounded conductor completely screens the interior electric field.  It follows that all multipole (traceless) moments of the deflector-and-shield charge distribution must exactly vanish, as any non-zero multipole moment would induce a power-law electric field at long distances.  Only \emph{trace} moments, which do not contribute to the electric field outside the shield, can be non-zero.  
Therefore, outside a shielded deflector, the leading contribution to $\rho_\x$ in Eq.~(\ref{eq:multipole0}) is the charge-radius piece of $\rho_\x^{(2)}$, 
$\frac{1}{6} \mathcal{R}_\defl^2 \lap^2 G(\xv)$, where 
\be
\label{eq:chargeR}
\mathcal{R}_\defl^2 \equiv \int d^3 \xv^\p ~ |\xv^\p|^2 \, \rho_\defl (\xv^\p).
\ee
For a given total charge $e q_\defl$ inside a spherical shield of radius $R$, the shield itself carries an opposite charge $-e q_\defl$, contributing $-e q_\defl R^2$ to the integral of Eq.~(\ref{eq:chargeR}).  If $\rho_\defl$ is positive-semidefinite, then $\mathcal{R}_\defl^2 \ge -e q_\defl R^2$, and this bound is saturated by concentrating the charge $q_\defl$ at the origin.  In this sense, our idealized geometry of a point charge at the origin, surrounded by a grounded shield of radius $R$, is the ``ideal charge-radius'' deflector geometry. 

From Eqs.~(\ref{eq:multipole0}), (\ref{eq:multipole}), and (\ref{eq:chargeR}), the induced DM charge density for this geometry can be computed for any DM velocity distribution.  For a shifted-Maxwellian, 
\be
\label{eq:fv}
f(\vv) \simeq \left( \frac{1}{\pi v_0^2} \right)^{3/2} e^{-|\vv - \vv_\text{wind}|^2/v_0^2},
\ee
the charge-radius contribution to $\rho_\x$ is
\be
\label{eq:rhoanalytic}
\rho_\x (\xv , t) \simeq \frac{2}{9} ~e^{i \omega t} ~  \rho_\x^{\text{Debye}}  \left( \frac{R}{|\xv|} \right)^3  \xi \, e^{- \xi^2} \left[ 2 \pi^{-1/2} c_\text{w} (1 - s_\text{w}^2 \xi^2) + e^{c_\text{w}^2 \xi^2} \xi \left( 2 c_\text{w}^2 (1 - s_\text{w}^2 \xi^2) - s_\text{w}^2\right) \text{erfc}(-c_\text{w} \xi)\right]
~,
\ee
where we have defined the wind-to-dispersion velocity ratio $\xi \equiv v_\text{wind} / v_0$, $c_\text{w} \equiv \cos{\theta_\text{wind}}$, $s_\text{w} \equiv \sin{\theta_\text{wind}}$, and $\theta_\text{wind}$ is the relative angle between $\xv$ and the DM wind, $\vv_\text{wind}$. We have also defined the Debye prefactor, analogous to Eq.~(\ref{eq:debye}),
\be
\label{eq:debye2}
\rho_\x^\text{Debye}  \equiv - \frac{(e \qeff)^2 \rhodm}{m_\x} ~ \frac{\langle \phi_\defl \rangle}{T}
~,~  \langle \phi_\defl \rangle = \frac{3}{8 \pi} \, \frac{e q_\defl}{R} 
~,
\ee
where $\langle \phi_\defl \rangle$ is the volume-averaged electric potential of the deflector. Note that Eq.~(\ref{eq:rhoanalytic}) vanishes in the slow-wind limit ($\xi \to 0$) since $f(v \, \hat{\vv}) \to f(v)$ becomes independent of $\xv^\p$ and hence $\lap^{2} G \to \lap^2 \frac{1}{|\xv|} = \delta^{(3)}(\xv)$ vanishes away from the origin (see Eqs.~(\ref{eq:Gdef}) and (\ref{eq:multipole})). For $|\xv| \gtrsim R$, we have checked that the leading order terms in the Taylor series of Eq.~(\ref{eq:multipole0}) (corresponding to the approximate expression in Eq.~(\ref{eq:rhoanalytic})) agree within $\order{10} \%$ with numerical evaluations of Eq.~(\ref{eq:rho5}).

As the earth rotates, $\theta_\text{wind}$ sweeps over a range of angles. The time-dependence of $\rho_\x$ over a sidereal day is shown in the left panel of Fig.~\ref{fig:angleandcurrent}, assuming that the deflector-detector axis is aligned with the direction of the DM wind at time $t = 0$~\cite{Griffin:2018bjn}. Evaluating the power spectral density of Eq.~(\ref{eq:rhoanalytic}), we find that most of the signal power peaks at $\sim \omega \pm n \, \omega_\oplus$ with $n=1$, with a large fraction of the remaining power in $n= 0,2$, where $\omega_\oplus$ is the frequency of a sidereal day. 

The far-field expansion of the induced DM current density ($\jx$) in Eq.~(\ref{eq:j1}) proceeds in a similar manner. Analogous to Eq.~(\ref{eq:multipole0}), we find 
\be
\label{eq:multipole2}
\jx (\xv, t) \simeq \frac{(e \qeff)^2 \rhodm}{m_\x^2} ~ e^{i \omega t} \, \left( \jv_E^{(2)} + \jv_E^{(3)} + \cdots + \jv_\rho^{(1)} + \jv_\rho^{(2)} +  \jv_\rho^{(3)} + \cdots \right)
~,
\ee
where
\begin{align}
\label{eq:multipole3}
\jv_E^{(2)} &\equiv  G_E(\xv) ~ \int d^3 \xv^\p ~ \E_\defl(\xv^\p)
\nl
\jv_E^{(3)} &\equiv - \lap_i G_E(\xv) ~ \int d^3 \xv^\p ~ x^{\p i } \,  \E_\defl(\xv^\p) 
\nl
\jv_\rho^{(1)} &\equiv \G_\rho(\xv) ~ Q_\defl
\nl
\jv_\rho^{(2)} &\equiv -\lap_i \G_\rho(\xv) ~p_\defl^i 
\nl
\jv_\rho^{(3)} &\equiv \frac{1}{6}  \lap_i \lap_j \G_\rho(\xv) \left( Q_\defl^{i j} +\delta^{ij} \mathcal{R}_\defl^2 \right)
~,
\end{align}
and
\be
\label{eq:Gdef2}
G_E (\yv) \equiv  \int dv ~ v ~\frac{f(v \, \hat{\yv})}{|\yv|^2}
~~,~~
\G_\rho (\yv) \equiv -  \int dv ~ v ~ \yv ~\frac{f(v \, \hat{\yv})}{|\yv|^2}
~.
\ee
As before, each $\jv_{E, \rho}^{(j)}$ term scales as $|\xv|^{-j}$.  For the same reasons as presented above, for a deflector surrounded by a grounded shield, the first $\jv_\rho^{(i)}$ term that contributes is the charge-radius piece of $\jv_\rho^{(3)}$. The presence of a grounded conductor also implies that $\jv_E^{(2)}$ vanishes in the quasi-static limit since $\E_\defl (\xv^\p) \simeq - \lap^\p \phi_\defl (\xv^\p)$ is a total derivative in the integrand of Eq.~(\ref{eq:multipole3}). Hence, $\jv_E^{(3)}$ and $\jv_\rho^{(3)}$ are the leading order terms in the far-field expansion of Eq.~(\ref{eq:multipole2}). 

For a deflector surrounded by a grounded shield, $\jv_E^{(3)}$ and $\jv_\rho^{(3)}$ are both proportional to the charge-radius-squared of the deflector, leading to
\be
\jv_E^{(3)} + \jv_\rho^{(3)} = - \frac{1}{6} \, \mathcal{R}_\defl^2 \int dv ~ v ~ \lap_i ~ \left( \xv ~ \lap_i \frac{f(v \, \hat{\xv})}{|\xv|^2} \right)
~,
\ee
where a sum over $i = 1,2,3$ is implied. Since the analytic far-field expression for $\jv_\x$ is more lengthy and cumbersome than $\rho_\x$, we do not present it explicitly here. However, we do note that similar to the charge density expression in Eq.~(\ref{eq:rhoanalytic}), $\jv_\x \propto v_\text{wind} \, (R/|\xv|)^3$. In the right panel of Fig.~\ref{fig:angleandcurrent}, we show these results for $\jv_\x$ projected on the $x-y$ plane and normalized by the dimensionful quantity,
\be
\label{eq:jdebye}
j_\x^\text{Debye} \equiv \rho_\x^\text{Debye} \, v_\text{wind}
~.
\ee
%

\section*{Electric Field Signal Calculation for Massive Dark Photons}

Here we show that for the electric field signal, the dark photon mass is irrelevant for distance scales smaller than $\mAp^{-1}$. In general, the hidden sector electric potential, $\phi^\p$, is sourced by SM and hidden sector charge densities ($\rho$ and $\rho^\p$, respectively), as dictated by its classical equation of motion. In the propagating mass-diagonal basis, this is
\be
 (\partial^2 + \mAp^2) \phi^\p =  \rho^\p + \eps \, \rho
~.
\ee
For a SM point charge, such as a stationary electron, this is easily solved via Green's functions such that
\be
\phi^\p = \frac{\eps \, e}{4 \pi} \, \frac{e^{-\mAp \, r}}{r}
~,
\ee
whereas $\phi^\p$ sourced by a hidden sector charge takes a similar form after making the replacement $\eps e \to e^\p$. The linear combination of fields that couples to SM particles is given by $\phi_\text{vis} \equiv \phi + \eps \, \phi^\p$ (see Eq.~(\ref{eq:visinvbasis}) below). From these simple examples, we see that the $\mAp \to 0$ limit is equivalent to considering length scales $r \ll \mAp^{-1}$. Furthermore, any calculation of the signal discussed in the main body of this text will involve combinations of $\eps$ and $e^\p$ in the form of the effective electromagnetic millicharge, as defined in Eq.~(\ref{eq:qeff}). In the next section, we illustrate this point in a more detailed calculation involving dynamical sources.

\section*{Magnetic Field Signal Calculation for Massive Dark Photons}

We now perform the proper calculation of the magnetic field signal for a massive dark photon. Any source that generates the oscillating electric field in the deflector region of Fig.~\ref{fig:setup} also creates a hidden electric field of strength, $\Ep_\defl \simeq \eps \E_\defl$, on length scales smaller than $\mAp^{-1}$, where $\eps$ ($\ll 1$) is the kinetic mixing parameter~\cite{Graham:2014sha}. Hence, for dark photons much lighter than the inverse geometric size of the deflector ($\mAp \ll \text{meter}^{-1} \sim 10^{-7} \ \eV$), DM with unit charge under the dark photon field is deflected, setting up a hidden sector current. A calculation nearly identical to that of Eqs.~(\ref{eq:debye0})-(\ref{eq:jestimate}) gives
\be
\label{eq:current2}
\jp \propto \eps \, e^{\p 2}
~,
\ee
where $e^\p$ is the dark photon gauge coupling. For simplicity, let us approximate $\jp$ as constant over a cylindrical conducting shield of radius $R_\text{sh}$. As we will show below, this toy example will demonstrate that calculations of visible magnetic fields sourced by a DM current density qualitatively reduce to the case of a massless dark photon when the dark photon is much lighter than the inverse length scale of the detection region.

Our task is to calculate how this oscillating hidden sector current density ($\jp$) sources visible fields ($\Evis$ and $\Bvis$) inside a conducting cavity. For the idealized setup, we imagine that the cylindrical conducting shield of radius, $R_\text{sh}$, is placed radially inside a cylindrical dark current of radius, $R_j$, such that $R_\text{sh} < R_j$. Moving radially outwards, we denote each region as follows:
\begin{itemize}
\label{eq:regions}
\item \textbf{Region 1}: $r < R_\text{sh}$ (inside the conducting shield)
\item \textbf{Region 2}: $R_\text{sh} < r < R_j$ (outside the conducting shield and inside the region of current density)
\item \textbf{Region 3}: $r > R_j$ (outside the region of current density, i.e., vacuum)
\end{itemize}
Above, $r$ is the cylindrical radial coordinate. We are interested in solving for the SM and hidden sector electric fields, $\E$ and $\Ep$, respectively. The corresponding magnetic fields can then be obtained by applying $\lap \times \E = - \partial_t \B$, and similarly for $\Ep$ and $\Bp$. Taking all time-dependences to be of the form $\sim e^{i \omega t}$, $\E$ and $\Ep$ are piecewise solutions to
\begin{align}
\label{eq:maxwell1}
(\lap^2 + \omega^2) \E &= 0
\nl
(\lap^2 + k^2) \Ep &= i \omega \jp ~ \Theta (R_j-r)
~,
\end{align}
applied in each radial region, where $\Theta$ is the Heaviside step-function and $k^2 \equiv \omega^2 - \mAp^2$. We take the current density, $\jp$,  to be aligned with the symmetry axis of the cylindrical conductor, i.e., along the $\hat{z}$ direction. In each region, we also define the alternate basis given by ``visible" and ``invisible" fields,
\be
\label{eq:visinvbasis}
\E_\text{vis} \equiv \E + \eps \Ep ~,~ \E_\text{inv} \equiv \Ep - \eps \E
~,
\ee
and similarly for $\B$ and $\B^\p$. The visible basis corresponds to the linear combination of fields that couples to SM electromagnetic currents in the presence of a kinetically mixed dark photon~\cite{Graham:2014sha}. The invisible linear combination is the orthogonal sterile mode.  

In each region, the general solution to Eq.~(\ref{eq:maxwell1}) is of the form
\be
\label{eq:besselsol}
E_{1,2,3} \sim c_J J_0 (\omega r) + c_Y Y_0 (\omega r) ~~,~~ E^\p_{1,2,3} \sim c_J^\p J_0 (k r) + c_Y^\p Y_0 (k r) + \order{j^\p}
~,
\ee
where the subscripts denote the different radial regions outlined above, $c_{J,Y}$ and $c_{J,Y}^\p$ are as of yet undetermined constants, and $J_n$ and $Y_n$ are Bessel functions of the first and second kind, respectively. Hence, there are four undetermined coefficients for each of the three regions, implying a total of 12 coefficients. These coefficients are fixed by imposing a total of 12 boundary conditions (BC). 
These are organized as:
\begin{enumerate}
\item \textbf{non-singular at origin} (2 BC)
\\ $E_1(0) = \text{finite} ~,~ E^\p_1(0) = \text{finite}$
\item \textbf{vanishing of the visible electric field at the conductor boundary} (2 BC)
\\ $E_{\text{vis}, 1} (R_\text{sh}) = E_{\text{vis}, 2} (R_\text{sh}) = 0$
\item \textbf{continuity of the invisible electric field and its derivative at the conductor boundary} (2 BC)
\\ $E_{\text{inv}, 1} (R_\text{sh}) = E_{\text{inv}, 2} (R_\text{sh}) ~,~ \partial_r E_{\text{inv}, 1} (R_\text{sh}) = \partial_r E_{\text{inv}, 2} (R_\text{sh})$ 
\item \textbf{continuity of electric fields and their derivatives at the current density boundary} (4 BC)
\\ $E_2 (R_j) = E_3 (R_j) ~,~ \partial_r E_2 (R_j) = \partial_r E_3 (R_j) ~,~ E^\p_2 (R_j) = E^\p_3 (R_j) ~,~ \partial_r E^\p_2 (R_j) = \partial_r E^\p_3 (R_j)$ 
\item \textbf{at distances far from the current density, electric fields propagate radially outwards} (2 BC)
\\ $E_3 \sim H_0^{(2)} (\omega r) ~,~ E_3^\p \sim H_0^{(2)} (k r)$
\end{enumerate}
making up a total of 12 boundary conditions. In condition 5, $H_n^{(2)}$ is a Hankel function of the second kind.

Conditions 1 and 5 are physical demands on the system. Condition 2 imposes that electrons do not experience forces along the surface of an ideal conductor. Condition 3 is tied to the fact that a conductor should have no effect on the invisible mode. 

To justify condition 4, note that integrating $\lap \times \Ep = - \partial_t \Bp$ around a small square loop straddling the interior and exterior of the current boundary implies that $E^\p$ is continuous across $r = R_j$. Since $E^\p = - \partial_t A^\p = - i \omega A^\p$, $A^\p$ is also continuous across this boundary. Maxwell's equation for a massive Proca field can be rewritten as $\lap \times \Bp = \jp + i \omega \Ep - \mAp^2 \boldsymbol{A}^\p$.
Note that $j^\p$ is discontinuous at the current boundary, but its radial integral is smooth. Hence, the fact that $\int d r \, j^\p$, $E^\p$, and $A^\p$  are all continuous, implies that $B^\p$ is continuous as well. Finally, $\lap \times \Ep = -i \omega \Bp$ implies that $\partial_r E^\p$ is also continuous. A similar argument holds for the SM electric field, $E$. This justifies condition 4.

After imposing all 12 boundary conditions on the piecewise solutions of Eq.~(\ref{eq:besselsol}), we find that the visible magnetic field inside the conducting shield is given by
\begin{align}
\label{eq:Bvis1}
\B_{\text{vis},1} &= \frac{i \pi \, \eps j^\p \, R_j \, e^{i \omega t} \boldsymbol{\hat{\phi}}}{2 \, k \, J_0(\omega R_\text{sh})} ~ \bigg( \omega \, J_1 (\omega r) \, \Big[ H_1^{(2)} (k R_j) \, \big( J_0(k R_\text{sh}) - J_0 (k R_j)\big) + J_1(k R_j) \,  H_0^{(2)} (k R_j) \Big]
\nl
&\qquad \qquad \qquad \quad ~ \,
 - k \, J_1(k r) \, H_1^{(2)} (k R_j) \, J_0(\omega R_\text{sh}) \bigg)
 ~.
\end{align}
In the $\mAp \to 0$ limit, this simplifies to
\be
\label{eq:Bvis2}
\B_{\text{vis}, 1} \simeq \frac{\eps \, j^\p \, J_1(\omega r)}{\omega \, J_0(\omega R_\text{sh})} 
\, e^{i \omega t} \, \boldsymbol{\hat{\phi}} \simeq \frac{1}{2} \, \eps \, j^\p \, r \, e^{i \omega t} \, \boldsymbol{\hat{\phi}}
~,
\ee
where in the second equality we took $\omega \, R_\text{sh} \ll 1$. Using Eqs.~(\ref{eq:qeff}) and (\ref{eq:current2}),  we see that the expression above is the standard result for a magnetic field sourced by a cylindrical current composed of truly charged particles. 

In the limit that $\mAp \gg \omega$, we make the replacement $k \to -i \kappa$, where $\kappa \equiv \sqrt{\mAp^2 - \omega^2} \, $. Expanding the Hankel functions around large $\mAp$, we then find that Eq.~(\ref{eq:Bvis1}) reduces to
\be
B_{\text{vis} , 1} \propto e^{- \mAp \, R_j}
~.
\ee
The exponential in the above expression indicates that the \emph{effective} millicharge limit corresponds to
\be
\mAp \ll R_j^{-1} 
~,
\ee
and that signals are exponentially suppressed for $\mAp \gg R_j^{-1}$.

\end{document}